\documentclass[letterpaper]{article}%
\usepackage{amsfonts}
\usepackage{amssymb}
\usepackage{amsmath}
\usepackage{graphicx}%
\providecommand{\U}[1]{\protect\rule{.1in}{.1in}}

\newtheorem{theorem}{Theorem}

\newtheorem{proposition}{Proposition}

\ifx\pdfoutput\relax\let\pdfoutput=\undefined\fi
\newcount\msipdfoutput
\ifx\pdfoutput\undefined\else
\ifcase\pdfoutput\else
\ifx\paperwidth\undefined\else
\ifdim\paperheight=0pt\relax\else\pdfpageheight\paperheight\fi
\ifdim\paperwidth=0pt\relax\else\pdfpagewidth\paperwidth\fi
\fi\fi\fi
\begin{document}

\title{Follow the Math!: \\The mathematics of quantum mechanics as the mathematics of set partitions
linearized to (Hilbert) vector spaces.}
\author{David Ellerman\\University of Ljubljana, Slovenia\\orcid.org/0000-0002-5718-618X\\david@ellerman.org}
\maketitle

Forthcoming in: \textit{Foundations of Physics}
\begin{abstract}
\noindent The purpose of this paper is to show that the mathematics of quantum
mechanics (QM) is the mathematics of set partitions (which specify
indefiniteness and definiteness) linearized to vector spaces, particularly in
Hilbert spaces. That is, the math of QM is the Hilbert space version of the math to describe objective indefiniteness that at the set level is the math of partitions. The key analytical concepts are definiteness versus
indefiniteness, distinctions versus indistinctions, and distinguishability
versus indistinguishability. The key machinery to go from indefinite to more
definite states is the partition join operation at the set level that
prefigures at the quantum level projective measurement as well as the
formation of maximally-definite state descriptions by Dirac's Complete Sets of Commuting Operators (CSCOs). This development
is measured quantitatively by logical entropy at the set level and by quantum
logical entropy at the quantum level. This follow-the-math approach supports
the Literal Interpretation of QM--as advocated by Abner Shimony among others
which sees a reality of objective indefiniteness that is quite different from
the common sense and classical view of reality as being ``definite all the way down."


\end{abstract}

\tableofcontents

\section{Introduction }

After a century of proliferating interpretations of quantum mechanics (QM), it
is time to ask the simple question: ``Where does the math of quantum mechanics
come from?".\footnote{The slogan ``Follow the money" means that finding the
source of an organization's or person's money may reveal their true nature. In
a similar sense, we use the slogan ``Follow the math!" to mean that finding
``where the mathematics of QM comes from" reveals a good deal about the key
concepts and machinery of the theory.} The purpose of this paper is to show
that the mathematics of quantum mechanics comes from the mathematics of set
partitions (the mathematical tools to describe indefiniteness and
definiteness) linearized to vector spaces, particularly Hilbert spaces.

Classical physics exemplified the common-sense idea that reality had definite
properties ``all the way down." At the logical level, i.e., Boolean subset
logic, each element in the Boolean universe set is either definitely in or not
in a subset, i.e., each element either definitely has or does not have a
property. Each element is characterized by a full set of properties, a view
that might be referred to as ``definite all the way down." This view of reality
was expressed by Leibniz's principle of the identity of indiscernible. If one
could always dig deeper into a definite reality to find attributes to
distinguish entities, then entities that were completely indistinguishable
would logically have to be identical. However, if there is no digging deeper
to find distinctions, then any remaining indefiniteness is objective.

It is now rather widely accepted that this common-sense always-definite view
of reality is not compatible with quantum mechanics (QM). If we think in terms
of only two positions, \textit{here} and \textit{there}, then in classical
physics a particle is either definitely \textit{here} or \textit{there}, while
in QM, the particle can be objectively ``neither definitely here nor there."
\cite[p. 144]{weinberg:dreams} Paul Feyerabend asserted that ``inherent
indefiniteness is a universal and objective property of matter." \cite[p.
202]{feyer:micro} This is not an epistemic or subjective indefiniteness of
location; it is an ontological or objective indefiniteness. The indefiniteness
or indistinguishability cannot be resolved by digging deeper with more
precision. The notion of \textit{objective indefiniteness} in QM has been most
emphasized by Abner Shimony.

\begin{quotation}
From these two basic ideas alone -- indefiniteness and the superposition
principle -- it should be clear already that quantum mechanics conflicts
sharply with common sense. If the quantum state of a system is a complete
description of the system, then a quantity that has an indefinite value in
that quantum state is objectively indefinite; its value is not merely unknown
by the scientist who seeks to describe the system. ...Classical physics did
not conflict with common sense in these fundamental ways.\cite[p.
47]{shim:reality}
\end{quotation}

Since our natural common sense view of the world is ``definite all the way
down" (as in classical physics), how can we describe an indefinite actuality?
The basic mathematical, indeed logical, concept that describes indefiniteness
and definiteness is the notion of a \textit{partition on a set}. It is the
purpose of this paper to show that the mathematics of quantum mechanics is
essentially the mathematics of partitions linearized to (Hilbert) vector
spaces. This substantiates that key analytical concepts in QM are
indefiniteness and definiteness, indistinction and distinction, and
indistinguishability and distinguishability. Key machinery of QM such as
projective measurement and the specification of maximally definite states by
Dirac's Complete Sets of Commuting Operators (CSCOs) will also be seen as the
linearization of the corresponding machinery in the logic of partitions.

\section{Partitions: The logical concept to describe indefiniteness and
definiteness}

Given a universe set $U=\left\{  u_{1},...,u_{n}\right\}  $, a
\textit{partition} $\pi=\left\{  B_{1},...,B_{m}\right\}  $ is a set of
subsets $B_{j}\subseteq U$ (for $j=1,...,m$) called \textit{blocks} that are disjoint
and whose union is $U$.\footnote{Since our purpose is conceptual clarity, not
mathematical generality, we will stick to the finite sets and dimensions
throughout.} A \textit{distinction} (or \textit{dit}) of a partition $\pi$ is
an ordered pair of elements $\left(  u_{i},u_{k}\right)  \in U\times U$ in
different blocks of $\pi$, and $\operatorname*{dit}\left(  \pi\right)
\subseteq U\times U$ is the set of all distinctions, called the
\textit{ditset} of $\pi$. An \textit{indistinction} (or \textit{indit}) of
$\pi$ is an ordered pair of elements in the same block of $U$, and the set of
all indistinctions $\operatorname*{indit}\left(  \pi\right)  \subseteq U\times
U$, called an \textit{indit set}, is the equivalence relation associated with
$\pi$ where the blocks are the equivalence classes. The ditset and indit set
of a partition are complements, i.e., they are disjoint and their union is
$U\times U$.

Each block $B_{j}\in\pi$ of a partition should be thought of as being
indefinite or indistinct between its elements $u_{i},u_{k}\in B_{j}$.
Partitions naturally arise as the inverse-images $f^{-1}=\left\{
f^{-1}\left(  y\right)  \right\}  _{y\in f\left(  U\right)  }$ of functions
$f:U\rightarrow Y$. In particular, a \textit{numerical attribute} is a
function $f:U\rightarrow\mathbb{R}$ into some set of values which we can take as the real numbers
$\mathbb{R}$. 
Each block $f^{-1}\left(  r\right)  $ in the partition $f^{-1}$ then
represents the constant set of all elements $u_{i}\in U$ taking the value
$f\left(  u_{i}\right)  =r\in
\mathbb{R}$. When the set $U$ is taken as the outcome set or sample space of a finite
probability distribution [with equiprobable points or point probabilities
$p_{i}=\Pr\left(  u_{i}\right)  $], then the numerical attribute is a random
variable. As an aid to intuition, these simple concepts at the logical level
might be seen as the elementary forms of the more developed mathematical
concepts of quantum mechanics as illustrated in Table 1. These connections
will be further developed in the later section on the Yoga of Linearization.

\begin{center}%
\begin{tabular}
[c]{|l|l|}\hline
Logical concept & QM concept\\\hline\hline
Block of a partition & Eigenspace\\\hline
Elements in a block & Eigenvectors with same eigenvalue\\\hline
Numerical attribute & Observable (self-adjoint operator)\\\hline
Elements of $U$ & Basis eigenvectors of an observable\\\hline
Values of attribute & Eigenvalues of an observable\\\hline
Partition on a set & Direct-sum decomposition of eigenspaces\\\hline Indistinctions & Coherences (density matrix non-zero entries\\\hline
\end{tabular}

Table 1: Partition logical precursors for QM math concepts
\end{center}

Subsets of a set and partitions on a set are mathematically dual concepts
\cite{law:sets}. The Boolean logic of subsets (usually presented in the
special case of ``propositional logic") thus has a dual mathematical logic, the
logic of partitions \cite{ell:lop}. The ``logical" concepts that prefigure the
mathematics of QM are those of the logic of partitions.

In the Boolean logic of subsets, the powerset $\wp\left(  U\right)  $ (all
subsets of $U$) forms a lattice where the partial order is set inclusion, the
join (least upper bound) and meet (greatest lower bound) are union and
intersection respectively, and the top and bottom of the lattice are the
universe set $U$ and the empty set $\emptyset$ respectively.

In the dual logic of partitions, the set $\Pi\left(  U\right)  $ of partitions
on $U$ also forms a lattice where the partial order is refinement. Given
another partition $\sigma=\left\{  C_{1},...,C_{m^{\prime}}\right\}  $ on $U$,
the partition $\sigma$ is \textit{refined by} $\pi$, written, $\sigma
\precsim\pi$, if for every block of $\pi$, there is a block of $\sigma$
containing it. Intuitively, the blocks of $\pi$ can be obtained by chopping up
the blocks of $\sigma$. If $\pi$ and $\sigma$ are the inverse images of random
variables $f:U\rightarrow%
\mathbb{R}
$ and $g:U\rightarrow%
\mathbb{R}
$ respectively, the $\sigma\precsim\pi$ means that the random variable $f$ is
\textit{sufficient} for $g$, i.e., the value of $f$ determines the value of
$g$.

The \textit{join} $\pi\vee\sigma$ (least upper bound in the refinement
ordering) is the partition of $U$ whose blocks are all the non-empty
intersections $B_{j}\cap C_{j^{\prime}}$. To form the \textit{meet} $\pi
\wedge\sigma$ (greatest lower bound in the refinement ordering), think of two
intersecting blocks $B_{j}$ and $C_{j^{\prime}}$ as two overlapping blobs of
mercury that unify to make a larger blob. Doing this for all overlapping
blocks, the blocks of the meet are the subsets of $U$ that are a union of
certain blocks of $\pi$ and simultaneously a union certain blocks of $\sigma$
(and are minimal in that respect). The top of the lattice of partitions
$\Pi\left(  U\right)  $ is the maximally distinguished \textit{discrete
partition} $\mathbf{1}_{U}=\left\{  \left\{  u_{1}\right\}  ,...,\left\{
u_{n}\right\}  \right\}  $ whose blocks are all the singletons of the elements
of $U$ and the bottom is the minimally distinguished \textit{indiscrete
partition} $\mathbf{0}_{U}=\left\{  U\right\}  $ (nicknamed the ``Blob") which
blobs all the elements together into one indefinite
``superposition."\footnote{Many of the older texts \cite{Birk:lt} presented the
``lattice of partitions" upside down, i.e., with the opposite partial order, so
the join and meet as well as the top and bottom were interchanged.} The join
operation is the only one we will need as it prefigures a projective
measurement in quantum mechanics.

Underlying the duality between subsets (e.g., images of functions) and
partitions (inverse-images of functions) is the duality between elements of
subsets and distinctions of a partitions, the `its and dits' duality. In the
Boolean lattice $\wp\left(  U\right)  $ of subsets, the partial order
$S\subseteq T$ for $S,T\in\wp\left(  U\right)  $ is the inclusion of elements.
In the lattice of partitions $\Pi\left(  U\right)  $, the refinement partial
order $\sigma\precsim\pi$ is just the inclusion of distinctions, i.e.,
$\sigma\precsim\pi$ if and only if (iff) $\operatorname*{dit}\left(
\sigma\right)  \subseteq\operatorname*{dit}\left(  \pi\right)  $. Moreover,
wherever the logical partial order holds, there is an induced logical map. If
$S\subseteq T$, then there is the canonical injection $S\rightarrowtail T$,
and if $\sigma\precsim\pi$, then there is the canonical surjection
$\pi\twoheadrightarrow\sigma$ that carries each block $B_{j}\in\pi$ to the
unique block $C_{k}\in\sigma$ such that $B_{j}\subseteq C_{k}$. The top $\mathbf{1}_{U}$ of the partition lattice includes all
possible distinctions, i.e., $\operatorname*{dit}\left(  \mathbf{1}%
_{U}\right)  =U\times U-\Delta$ [where $\Delta$ is the diagonal of self-pairs
$\left(  u_{i},u_{i}\right)  $], just as the top $U$ of the the subset lattice
thus includes all possible elements. The bottom $\mathbf{0}_{U}$ of the
partition lattice has no distinctions, i.e., $\operatorname*{dit}\left(
\mathbf{0}_{U}\right)  =\emptyset$, just as the bottom $\emptyset$ of the
subset lattice has no elements. The ditset of the join in the partition
lattice is the union of the distinctions just as the join in the lattice of
subsets is union of the elements of the subsets. This duality between elements
and distinctions is illustrated in Table 2.

\begin{center}%
\begin{tabular}
[c]{|c||c|c|}\hline
{\small Dualities} & {\small Subset logic} & {\small Partition logic}%
\\\hline\hline
{\small Its or dits} & {\small Elements }$u${\small \ of }$S$ &
{\small Distinctions }$\left(  u,u^{\prime}\right)  ${\small \ of }$\pi
$\\\hline
{\small Partial order} & {\small Inclusion }$S\subseteq T$ & $\sigma
\precsim\pi${\small , }$\operatorname*{dit}\left(  \sigma\right)
\subseteq\operatorname*{dit}\left(  \pi\right)  $\\\hline
{\small Logical maps} & ${ S\rightarrowtail T}$ & ${ \pi
\twoheadrightarrow\sigma}$\\\hline
All & All elements $U$ & All distinctions $\mathbf{1}_{U}$\\\hline
None & No elements $\emptyset$ & No distinctions $\mathbf{0}_{U}$\\\hline
Join & $S\cup T$ & $\operatorname*{dit}\left(  \pi\vee\sigma\right)
=\operatorname*{dit}\left(  \pi\right)  \cup\operatorname*{dit}\left(
\sigma\right)  $\\\hline
\end{tabular}

Table 2: Its \& Dits duality
\end{center}

Following Heisenberg,\footnote{In his sympathetic interpretation of
Aristotle's treatment of substance and form, Heisenberg refers to the
substance as: ``a kind of indefinite corporeal substratum, embodying the
possibility of passing over into actuality by means of the form."\cite[p.
148]{heisen:phy-phil} Heisenberg's ``potentiality" ``passing over into actuality
by means of the form" should be seen as the actual indefinite ``passing over
into" the actual definite by being objectively in-formed through the making of
distinctions.} we might express this duality by going back to the ancient
Greek metaphysical notions of substance (or matter) and form
\cite{ains:form-matter}. At this simple level, one can still discern two
`creation stories' corresponding to the classical (definite all the way down)
and the quantum (objective indefiniteness) versions. These two stories can be
represented by moving from the bottom up the two logical lattices illustrated
in Figure 1 where the universe now consists three states $U=\left\{
a,b,c\right\}  $.%

\begin{figure}[h!]
	\centering
	\includegraphics[width=0.7\linewidth]{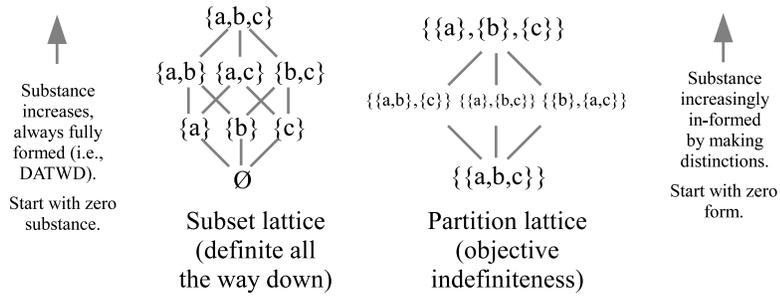}
	\caption{The two creation stories illustrated by the dual lattices}
	\label{fig:fig1twolattices}
\end{figure}

Subset creation story: In the Beginning was the Void (no substance) and
then fully definite elements ("Its") were created until the universe $U$ was created.

Partition creation story: In the Beginning was the Blob--all substance (energy)
with no form--and then, in a Big Bang, distinctions ("Dits") were created as the substance was increasingly in-formed to
reach the universe $U$.

\section{Intuitive imagery for superposition}

How should one imagine a quantum superposition? The most misleading imagery in
QM is the \textit{classical }interpretation of superposition (Figure 2) as the
addition of two definite waves to get another \textit{definite} wave.%

\begin{figure}[h!]
	\centering
	\includegraphics[width=0.7\linewidth]{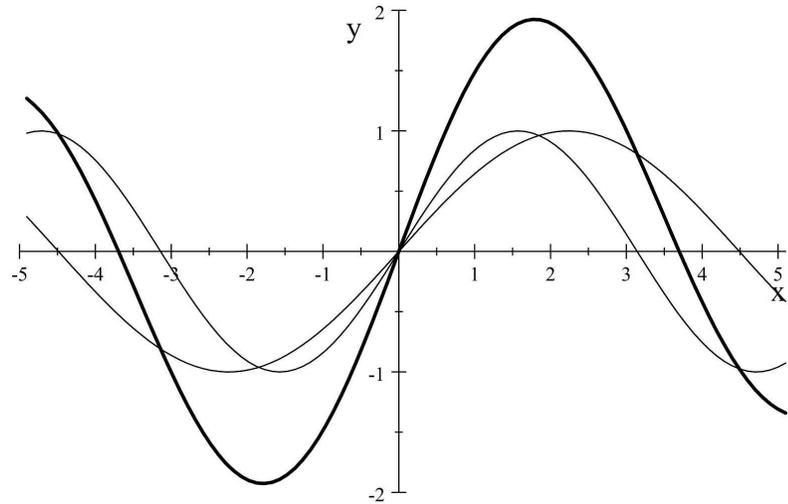}
	\caption{Imagery for \textit{classical} superposition}
	\label{fig:fig2wavesuperposition}
\end{figure}

But there seem to be no actual physical waves in QM, much to the
disappointment of Schr\"{o}dinger and others. The complex numbers are the
natural mathematics to describe waves so the misleading wave formalism is always there.

\begin{quotation}
\noindent Such analogies have led to the name `Wave Mechanics' being sometimes
given to quantum mechanics. It is important to remember, however, that the
superposition that occurs in quantum mechanics is of an essentially different
nature from any occurring in the classical theory, as is shown by the fact
that the quantum superposition principle demands indeterminacy in the results
of observations in order to be capable of a sensible physical interpretation.
The analogies are thus liable to be misleading. \cite[p. 14]{dirac:principles}
\end{quotation}

The complex numbers are needed in the mathematics of QM (among other reasons)
since they are the algebraically-complete extension of the reals so the
real-valued observables will have a full set of eigenvectors, \textit{not}
because there are any physical waves.

The quantum (as opposed to classical) interpretation of superposition is the
addition of two definite states to get a new state \textit{indefinite}
\textit{between the definite states}--not the `double-exposure' image (e.g.,
not being simultaneously \textit{here} and \textit{there} as in so much of the
popular science literature) suggested by the wave interpretation.\footnote{The
indefiniteness interpretation of qubit $a\left\vert 0\right\rangle
+b\left\vert 1\right\rangle $ is more routine in quantum information and
computation theory \cite{nielsen-chuang:bible} as opposed to the being
simultaneously $\left\vert 0\right\rangle $ and $\left\vert 1\right\rangle $
version of superposition.} In Figure 3, the superposition of the two definite
isosceles triangles is the indefinite triangle which is indefinite on where
the two definite triangles are distinct (the labeling of the equal sides) and
is definite on where the two triangles do not differ (the $aA$%
-axis).\footnote{There is at least an analogy between superposition in QM and
abstraction in mathematics. In Frege's example of a set of parallel directed
line segments oriented in the same way, the abstraction ``direction" is
definite on what is common between the lines and indefinite on how they differ
\cite{ell:abstr}. In QM, the emphasis is on the indefiniteness between the
definite eigenstates (glass half-empty) while in abstraction, the emphasis is
on the common definiteness between the instances (glass half-full).}%

\begin{figure}[h!]
	\centering
	\includegraphics[width=0.7\linewidth]{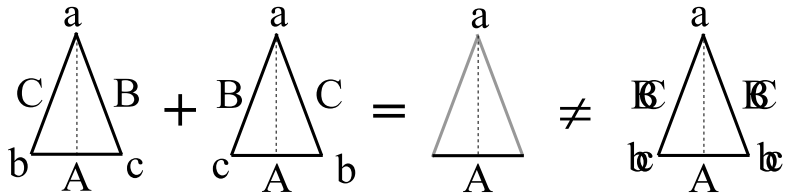}
	\caption{Imagery for \textit{Quantum} Superposition}
	\label{fig:fig3imageqsuperposition}
\end{figure}

\noindent The intuitions in the theater of our minds have evolved to think in
terms of a macroscopic spatial world, so one should not expect to have fully
definite classical imagery for indefinite quantum states. The best to expect
is probably a set of image `crutches' to illustrate one aspect or another as
in Figure 3.

At the simplest logical level, the pure and mixed quantum states for a single
particle can be illustrated by partitions as in Figure 4. There are four
possible eigenstates for a particle represented by $a,b,c,$ and $d$. The pure
state of all those eigenstates superposed is represented by the indiscrete
partition $\left\{  \left\{  a,b,c,d\right\}  \right\}  $ (written in
shorthand as $\left\{  abcd\right\}  $) and then distinctions are made (i.e.,
`measurements' are made) to get the other partition representations of the
mixed states of `orthogonal' (disjoint) superpositions such as $\left\{
\left\{  a,c\right\}  ,\left\{  b,d\right\}  \right\}  $ (or in shorthand
$\left\{  ac,bd\right\}  $).%

\begin{figure}[h!]
	\centering
	\includegraphics[width=0.7\linewidth]{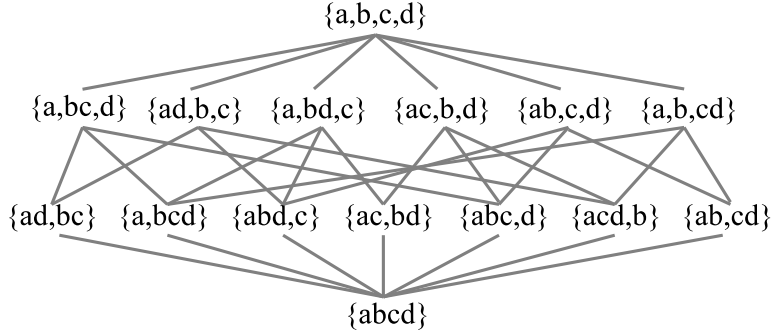}
	\caption{Lattice of partitions on $U=\left\{ a,b,c,d\right\} $}
	\label{fig:fig4latticeofpartitions4elements}
\end{figure}

\noindent In terms of sets (as in Figure 4), the superposition of the
eigenstates $\left\{  a\right\}  $ and $\left\{  c\right\}  $ is the state
$\left\{  a,c\right\}  $, which is the state indefinite between $\left\{
a\right\}  $ and $\left\{  c\right\}  $; not definitely $\left\{  a\right\}  $
and not definitely $\left\{  c\right\}  $, but definitely not $\left\{
b\right\}  $ or $\left\{  d\right\}  $. A distinction between $\left\{
a\right\}  $ and $\left\{  c\right\}  $, e.g., in the join $\left\{
ac,bd\right\}  \vee\left\{  a.bcd\right\}  =\left\{  a,bd,c\right\}  $, would
reduce the superposition $\left\{  a,c\right\}  $ to a mixture of the
eigenstates $\left\{  a\right\}  $ and $\left\{  c\right\}  $, e.g., in the
mixed state $\left\{  a,bd,c\right\}  $. In contrast to the misleading
superposition of waves imagery, these simple examples illustrate that the
superposition $\left\{  a,c\right\}  $ is not a definite state but is an
indefinite state that could reduce (e.g., with distinctions supplied by
$\left\{  a.bcd\right\}  $) to a mixture of the definite eigenstates $\left\{
a\right\}  $ or $\left\{  c\right\}  $.

\section{Logical entropy: the measure of distinctions}

Since partitions are the mathematical concept to represent distinctions and
indistinctions (or definiteness and indefiniteness), there should be a
quantitative measure (in the sense of measure theory) to quantify
distinctions. Since $U$ is a finite set and the set of distinctions of a
partition $\pi$ is finite, the obvious notion to measure distinctions is
simply the cardinality of the set of distinctions $\operatorname*{dit}\left(
\pi\right)  \subseteq U\times U$ normalized by the size of $U\times U$. Hence
the \textit{logical entropy of a partition} $\pi=\left\{  B_{1},...,B_{m}%
\right\}  $ for equiprobable points in $U$ is:

\begin{center}
$h\left(  \pi\right)  =\frac{\left\vert \operatorname*{dit}\left(  \pi\right)
\right\vert }{\left\vert U\times U\right\vert }=\frac{\left\vert U\times
U\right\vert -\left\vert \cup_{j}B_{j}\times B_{j}\right\vert }{\left\vert
U\times U\right\vert }=1-\sum_{j}\frac{|B_{j}|^{2}}{\left\vert U\right\vert
^{2}}=1-\sum_{j}\Pr\left(  B_{j}\right)  ^{2}$
\end{center}

\noindent where $\Pr\left(  B_{j}\right)  =\left\vert B_{j}\right\vert
/\left\vert U\right\vert $ is the probability that a random draw from $U$
gives an element of $B_{j}$ (\cite{ell:counting}, \cite{ell:nf4it},
\cite{ell:4open}). When the points of $U$ have the probabilities $p_{i}$ for
$i=1,...,n$, then:

\begin{center}
$h\left(  \pi\right)  =1-\sum_{j}\Pr\left(  B_{j}\right)  ^{2}$
\end{center}

\noindent where $\Pr\left(  B_{j}\right)  =\sum_{u_{i}\in B_{j}}p_{i}$. The
logical entropy of $\pi$ has an immediate interpretation; it is the
probability that in two independent random draws from $U$, one will obtain a
distinction of $\pi$--just as the probability of a subset $S\subseteq U$,
$\Pr\left(  S\right)  $ is the probability that in one random draw from $U$,
one will obtain an element of $S$. Since the indiscrete partition makes no
distinctions, its logical entropy is zero, $h\left(  \mathbf{0}_{U}\right)
=0$. The discrete partition makes all possible distinctions so its logical
entropy is $h\left(  \mathbf{1}_{U}\right)  =1-\sum_{i=1}^{n}p_{i}^{2}$,
which, in the equiprobable case, is $1-\frac{1}{n}$, the probability that the
second draw is not the same as the first draw.

This definition of logical entropy fulfills a program of Gian-Carlo Rota that
begins with the idea: ``The lattice of partitions plays for information the
role that the Boolean algebra of subsets plays for size or probability."
\cite[p. 30]{kung:rota} In Rota's Fubini Lectures (and in his lectures as
MIT), he argued that since partitions are dual to subsets, then
quantitatively, information is to partitions as probability is to subsets:

\begin{center}
$\frac{\text{Information}}{\text{Partitions}}\approx\frac{\text{Probability}%
}{\text{Subsets}}$.
\end{center}

\noindent Since ``Probability is a measure on the Boolean algebra of events"
that gives quantitatively the ``intuitive idea of the size of a set", we may
ask by ``analogy" for some measure ``which will capture some property that will
turn out to be for [partitions] what size is to a set." He then asks:

\begin{quote}
How shall we be led to such a property? We have already an inkling of what it
should be: it should be a measure of information provided by a random
variable. Is there a candidate for the measure of the amount of information?
\cite[p. 67]{rota:fubini}
\end{quote}

\noindent The underlying duality of elements and distinctions answers that
question. The lattice of partitions is isomorphic to the lattice of ditsets
partially ordered by inclusion (since refinement is just inclusion of
ditsets), and the normalized size of subsets and ditsets (equiprobable case)
gives the notions of probability $\Pr(S)=\frac{\left\vert S\right\vert
}{\left\vert U\right\vert }$ and logical entropy $h\left(  \pi\right)
=\frac{\left\vert \operatorname*{dit}\left(  \pi\right)  \right\vert
}{\left\vert U\times U\right\vert }$--as summarized in Table 3.

\begin{center}%
\begin{tabular}
[c]{|c||c|c|}\hline
& Logical Probability Theory & Logical Information Theory\\\hline\hline
`Outcomes' & Elements $u\in U$ finite & Dits $\left(  u,u^{\prime}\right)  \in
U\times U$ finite\\\hline
`Events' & Subsets $S\subseteq U$ & Ditsets $\operatorname*{dit}\left(
\pi\right)  \subseteq U\times U$\\\hline
Equiprobable points & $\ \Pr\left(  S\right)  =\frac{|S|}{\left\vert
U\right\vert }$ & $h\left(  \pi\right)  =\frac{\left\vert \operatorname*{dit}%
\left(  \pi\right)  \right\vert }{\left\vert U\times U\right\vert }$\\\hline
Point probabilities & $\ \Pr\left(  S\right)  =\sum\left\{  p_{j}:u_{j}\in
S\right\}  $ & $h\left(  \pi\right)  =\sum\left\{  p_{j}p_{k}:\left(
u_{j},u_{k}\right)  \in\operatorname*{dit}\left(  \pi\right)  \right\}
$\\\hline
Interpretation & $\Pr(S)=$ one-draw prob. of $S$-element & $h\left(
\pi\right)  =$ two-draw prob. of $\pi$-distinction\\\hline
\end{tabular}

Table 3: Classical logical probability theory and `classical' logical
information theory.
\end{center}

When the point probabilities on $U$ are given by the probability distribution
$p=\left(  p_{1},...,p_{n}\right)  $, then the logical entropy $h\left(
\pi\right)  $ is the product probability measure $p\times p$ (defined on
$U\times U$) of the ditset $\operatorname*{dit}\left(  \pi\right)  \subseteq
U\times U$. Logical entropy is the measure of information-as-distinctions.
Since the logical entropy is the value of a measure in the sense of measure
theory (unlike Shannon entropy \cite{ell:nf4it}), the compound notions of
logical entropy are naturally defined in the usual Venn diagram manner as
illustrated in Figure 5 which includes the conditional logical entropy
$h(\sigma|\pi)$ (the measure of the distinctions in $\sigma$ that were not in
$\pi$) and the mutual logical information $m\left(  \pi,\sigma\right)  $ (the
measure of the distinctions common to $\pi$ and $\sigma$).%

\begin{figure}[h!]
	\centering
	\includegraphics[width=0.5\linewidth]{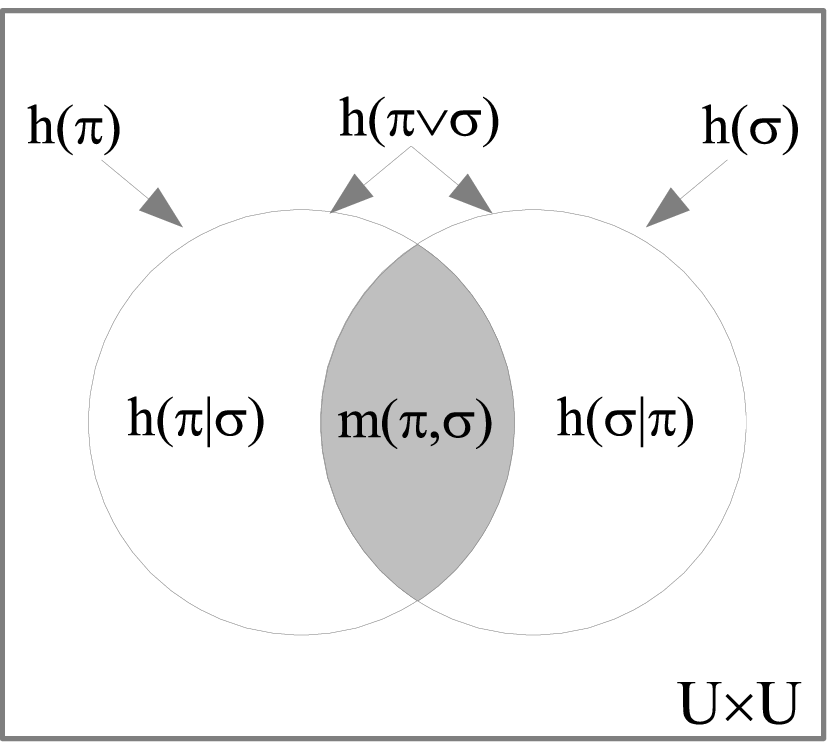}
	\caption{$h\left( \pi\vee\sigma\right) =h\left( \pi\right) +h\left( \sigma\right) -m\left( \pi,\sigma\right) =h\left( \pi|\sigma\right) +h\left( \sigma|\pi\right) +m\left( \pi,\sigma\right) $}
	\label{fig:fig5logentvennpartitions}
\end{figure}

\noindent The compound notion of logical entropy that we will make later use
of in the analysis of quantum measurement is the logical entropy $h\left(
\pi\vee\sigma\right)  $ of the join $\pi\vee\sigma$ which is the probability
measure $p\times p$ on $\operatorname*{dit}\left(  \pi\vee\sigma\right)
=\operatorname*{dit}\left(  \pi\right)  \cup\operatorname*{dit}\left(
\sigma\right)  $.

\section{Formulation using density matrices}

In quantum mechanics, the state of a system can be represented by state
vectors or by density matrices \cite[p. 102]{nielsen-chuang:bible}. The best
form for our purposes is density matrices because the relevant machinery
developed above about partitions and logical entropy can be reformulated using
`classical' density matrices over the reals.

Given a partition $\pi=\left\{  B_{1},...,B_{m}\right\}  $ on $U=\left\{
u_{1},...,u_{n}\right\}  $ with point probabilities $p=\left(  p_{1}%
,...,p_{n}\right)  $, an $n\times n$ density matrix $\rho\left(  B_{j}\right)
$ can be defined for each block $B_{j}\in\pi$ as follows:

\begin{center}
$\rho\left(  B_{j}\right)  _{ik}=\left\{
\begin{array}
[c]{l}%
\frac{\sqrt{p_{i}p_{k}}}{\Pr\left(  B_{j}\right)  }\text{ if }\left(
u_{i},u_{k}\right)  \in B_{j}\times B_{j}\\
0\text{ otherwise}%
\end{array}
\right.  $.
\end{center}

\noindent where $\Pr\left(  B_{j}\right)  =\sum_{u_{i}\in B_{j}}p_{i}$. Then
these density matrices for the blocks are combined to form the density matrix
$\rho\left(  \pi\right)  $ representing the partition $\pi$:

\begin{center}
$\rho\left(  \pi\right)  =\sum_{j=1}^{m}\Pr\left(  B_{j}\right)  \rho\left(
B_{j}\right)  $
\end{center}

\noindent so the entries are:

\begin{center}
$\rho\left(  \pi\right)  _{ik}=\left\{
\begin{array}
[c]{l}%
\sqrt{p_{i}p_{k}}\text{ if }\left(  u_{i},u_{k}\right)  \in
\operatorname*{indit}\left(  \pi\right) \\
0\text{ otherwise}%
\end{array}
\right.  $.
\end{center}

\noindent These density matrices over the reals are symmetric and have trace
(sum of diagonal elements) equal to $1$ since the diagonal elements are
$\sqrt{p_{i}p_{i}}=p_{i}$ for $i=1,...,n$. The probability $p_{i}$ of an
element $u_{i}$ is recovered as $\operatorname*{tr}\left[  P_{u_{i}}%
\rho\left(  \pi\right)  \right]  =\sqrt{p\,_{i}}\sqrt{p_{i}}=p_{i}$ [where
$P_{u_{i}}$ is the diagonal projection matrix with entries $\chi_{\left\{
u_{i}\right\}  }()$] which is the set version of the Born Rule. Assuming only
non-zero probabilities, the non-zero off-diagonal elements indicate the
indistinctions of $\pi$ where elements $u_{i}$ and $u_{k}$ `cohere' together
in the same block of the partition $\pi$ and are called ``coherences" in the
case of quantum density matrices (\cite[p. 303]{cohen-tennoudji:vol1-2};
\cite[p. 177]{auletta:qm}). Thus at the logical level, indistinctions
prefigure quantum coherences.

In the formula for logical entropy $h\left(  \pi\right)  =1-\sum_{j}\Pr\left(
B_{j}\right)  ^{2}$, the density matrix replaces the block probabilities and
the trace replaces the summation to give the same result:

\begin{center}
$h\left(  \pi\right)  =1-\sum_{j}\Pr\left(  B_{j}\right)  ^{2}%
=1-\operatorname*{tr}\left[  \rho\left(  \pi\right)  ^{2}\right]  $.
\end{center}

\noindent These real-valued density matrices encapsulate the `classical'
treatment of the mathematics of partitions that prefigures the quantum
treatment. Two of these classical results translate directly into the
corresponding results in quantum mechanics.

The first result is that the projective measurement in QM is classically just
the join of partitions. We start with the partition $\pi$ expressed by the
density matrix $\rho\left(  \pi\right)  $ and then we think of the partition
$\sigma=\left\{  C_{1},...,C_{m^{\prime}}\right\}  $ on $U$ as being the
inverse-image of a numerical attribute or `observable' $g:U\rightarrow%
\mathbb{R}
$. In QM, the effect of a projective measurement on a density matrix $\rho$ is
given by the \textit{L\"{u}ders mixture operation} (\cite[p. 279]{auletta:qm};
\cite{luders:meas}). For each block $C_{j}\in\sigma$, let $P_{C_{j}}$ be the
projection matrix that is a diagonal matrix with the diagonal elements given
by the characteristic function $\chi_{C_{j}}:U\rightarrow2=\left\{
0,1\right\}  $ of $C_{j}$. Then the L\"{u}ders mixture operation transforms
the density matrix $\rho\left(  \pi\right)  $ into the density matrix
$\hat{\rho}\left(  \pi\right)  $ according to the formula:

\begin{center}
$\hat{\rho}\left(  \pi\right)  =\sum_{C_{j}\in\sigma}P_{C_{j}}\rho\left(
\pi\right)  P_{C_{j}}$.

Classical L\"{u}ders Mixture Operation
\end{center}

\begin{theorem}
$\hat{\rho}\left(  \pi\right)  =\rho\left(  \pi\vee\sigma\right)  $.
\end{theorem}

\noindent Proof: A nonzero entry in $\rho\left(  \pi\right)  $ has the form
$\rho\left(  \pi\right)  _{ik}=\sqrt{p_{i}p_{k}}$ iff there is some block
$B\in\pi$ such that $\left(  u_{i},u_{k}\right)  \in B\times B$, i.e., if
$u_{i},u_{k}\in B$. The matrix operation $P_{C_{j}}\rho\left(  \pi\right)  $
will preserve the entry $\sqrt{p_{i}p_{k}}$ if $u_{i}\in C_{j}$, otherwise the
entry is zeroed. And if the entry was preserved, then the further matrix
operation $\left(  P_{C_{j}}\rho\left(  \pi\right)  \right)  P_{C_{j}}$ will
preserve the entry $\sqrt{p_{i}p_{k}}$ if $u_{k}\in C_{j}$, otherwise it is
zeroed. Hence the entries $\sqrt{p_{i}p_{k}}$ in $\rho\left(  \pi\right)  $
that are preserved in $P_{C_{j}}\rho\left(  \pi\right)  P_{C_{j}}$ are the
entries where both $u_{i},u_{k}\in B$ for some $B\in\pi$ and $u_{i},u_{k}\in
C_{j}$. Recall that $\operatorname*{dit}\left(  \pi\vee\sigma\right)
=\operatorname*{dit}\left(  \pi\right)  \cup\operatorname*{dit}\left(
\sigma\right)  $ so $\operatorname*{indit}\left(  \pi\vee\sigma\right)
=\operatorname*{indit}\left(  \pi\right)  \cap\operatorname*{indit}\left(
\sigma\right)  $--since the join of partitions is just the partition
corresponding to the equivalence relation resulting from intersecting two
equivalence relations (indit sets). These are the entries in $\rho\left(
\pi\vee\sigma\right)  $ corresponding to the blocks $B\cap C_{j}$ for some
$B\in\pi$, so summing over $C_{j}\in\sigma$ gives the result: $\sum_{C_{j}%
\in\sigma}P_{C_{j}}\rho\left(  \pi\right)  P_{C_{j}}=\hat{\rho}\left(
\pi\right)  =\rho\left(  \pi\vee\sigma\right)  $. $\square$

Our theme is that the vector space mathematics of QM is prefigured at the
logical level by the mathematics of partitions on sets. The above Theorem
shows that the standard partition operation of join is essentially the set
version of the projective measurement operation in QM. Note that partitions
have two separate roles in this set-based example; $\pi$ represents the state
being measured and $\sigma$ represents the numerical attribute (or observable)
being measured on that state. The join operation creates more distinctions
since $\operatorname*{dit}\left(  \pi\vee\sigma\right)  =\operatorname*{dit}%
\left(  \pi\right)  \cup\operatorname*{dit}\left(  \sigma\right)  $. The
off-diagonal non-zero entries in the density matrices represent
indistinctions, so the distinctions that are created by joining $\sigma$ with
$\pi$ will be indicated by those non-zero entries in $\rho\left(  \pi\right)
$ that are zeroed in $\hat{\rho}\left(  \pi\right)  =\rho\left(  \pi\vee
\sigma\right)  $. Logical entropy measures information-as-distinctions so the
non-zero off-diagonal entries that are zeroed, the indistinctions that become
distinctions (i.e., the coherences that are decohered in the quantum case),
will be measured by the increase in logical entropy.

\begin{theorem} [Set version of Measuring Measurement Theorem]
The sum of all the squares $p_{i}p_{k}$ of all the entries $\sqrt{p_{i}p_{k}}$
that were zeroed in the L\"{u}ders mixture operation that transforms
$\rho\left(  \pi\right)  $ into $\hat{\rho}\left(  \pi\right)  =\sum_{C_{j}%
\in\sigma}P_{C_{j}}\rho\left(  \pi\right)  P_{C_{j}}=\rho\left(  \pi\vee
\sigma\right)  $ is $h\left(  \pi\vee\sigma\right)  -h\left(  \pi\right)
=h\left(  \sigma|\pi\right)  $.
\end{theorem}

\noindent Proof: All the entries $\sqrt{p_{i}p_{k}}$ that got zeroed were for
ordered pairs $\left(  u_{i},u_{k}\right)  $ that were indits of $\pi$ but not
indits of $\pi\vee\sigma$, i.e., $\left(  u_{i},u_{k}\right)  \in
\operatorname{indit}\left(  \pi\right)  \cap\operatorname{indit}\left(
\pi\vee\sigma\right)  ^{c}=\operatorname{dit}\left(  \pi\right)  ^{c}%
\cap\operatorname{dit}\left(  \pi\vee\sigma\right)  =\operatorname{dit}\left(
\pi\vee\sigma\right)  -\operatorname{dit}\left(  \pi\right)  $. The sum of
products $p_{i}p_{k}$ for those pairs $\left(  u_{i},u_{k}\right)  $ is just
the product probability measure on that set $\operatorname{dit}\left(  \pi
\vee\sigma\right)  -\operatorname{dit}\left(  \pi\right)  $ which is $h\left(
\pi\vee\sigma|\pi\right)  $. And since $\operatorname{dit}\left(  \pi\right)
\subseteq\operatorname{dit}\left(  \pi\vee\sigma\right)  $, the measure on
$\operatorname{dit}\left(  \pi\vee\sigma\right)  -\operatorname{dit}\left(
\pi\right)  $ is $h\left(  \pi\vee\sigma|\pi\right)  =h\left(  \pi\vee
\sigma\right)  -h\left(  \pi\right)  =h\left(  \sigma|\pi\right)  $ (see
Figure 5) which is the information-as-distinctions that $\sigma$ added to the
information in $\pi$. $\square$

\textbf{Example}: If the four elements of $U=\left\{  a,b,c,d\right\}  $ were
equiprobable, the real-valued density matrix of the partition $\left\{
abc,d\right\}  $ is:

\begin{center}
$\rho\left(  \left\{  abc,d\right\}  \right)  =%
\begin{bmatrix}
\frac{1}{4} & \frac{1}{4} & \frac{1}{4} & 0\\
\frac{1}{4} & \frac{1}{4} & \frac{1}{4} & 0\\
\frac{1}{4} & \frac{1}{4} & \frac{1}{4} & 0\\
0 & 0 & 0 & \frac{1}{4}%
\end{bmatrix}
$.
\end{center}

\noindent The main partition operation representing going from an indefinite
state or partition to a more definite (i.e., more refined) one is the
\textit{join} operation, for instance: $\left\{  ac,bd\right\}  \vee\left\{
abc,d\right\}  =\left\{  ac,b,d\right\}  $ as in Figure 6:%

\begin{figure}[h!]
	\centering
	\includegraphics[width=0.7\linewidth]{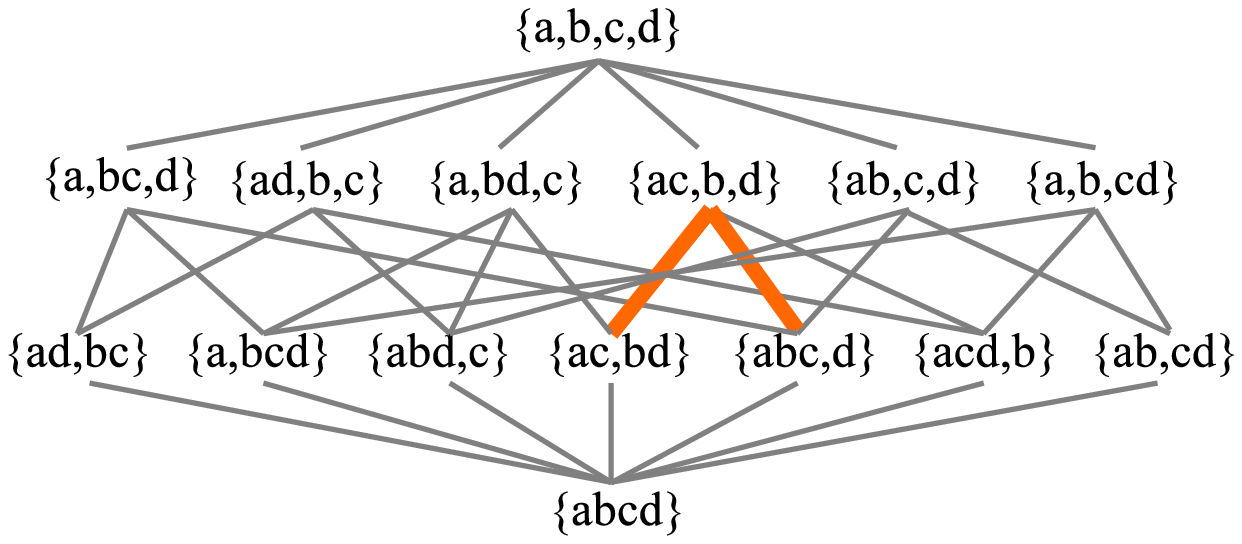}
	\caption{Partition join: $\left\{  ac,bd\right\}  \vee\left\{  abc,d\right\}
		=\left\{  ac,b,d\right\}  $}
	\label{fig:fig6partitionjoinop}
\end{figure}

\noindent The QM-version of the partition join is a projective measurement
described by the L\"{u}ders mixture operation. Since $\left\{  ac,bd\right\}
$ is being joined to $\left\{  abc,d\right\}  $, we need the projection
matrices to $\left\{  a,c\right\}  $ and to $\left\{  b,d\right\}  $ which are:

\begin{center}
$P_{\left\{  a,c\right\}  }=%
\begin{bmatrix}
1 & 0 & 0 & 0\\
0 & 0 & 0 & 0\\
0 & 0 & 1 & 0\\
0 & 0 & 0 & 0
\end{bmatrix}
$ and $P_{\left\{  b,d\right\}  }=%
\begin{bmatrix}
0 & 0 & 0 & 0\\
0 & 1 & 0 & 0\\
0 & 0 & 0 & 0\\
0 & 0 & 0 & 1
\end{bmatrix}
$.
\end{center}

\noindent The L\"{u}ders mixture operation pre- and post-multiplies the
pre-measurement density matrix $\rho\left(  \left\{  abc,d\right\}  \right)  $
by these two projection matrices and the result sums to yield the
post-measurement density matrix $\hat{\rho}$:

\begin{center}
$\hat{\rho}=P_{\left\{  a,c\right\}  }\rho\left(  \left\{  abc,d\right\}
\right)  P_{\left\{  a,c\right\}  }+P_{\left\{  b,d\right\}  }\rho\left(
\left\{  abc,d\right\}  \right)  P_{\left\{  b,d\right\}  }$

$%
\begin{bmatrix}
1 & 0 & 0 & 0\\
0 & 0 & 0 & 0\\
0 & 0 & 1 & 0\\
0 & 0 & 0 & 0
\end{bmatrix}%
\begin{bmatrix}
\frac{1}{4} & \frac{1}{4} & \frac{1}{4} & 0\\
\frac{1}{4} & \frac{1}{4} & \frac{1}{4} & 0\\
\frac{1}{4} & \frac{1}{4} & \frac{1}{4} & 0\\
0 & 0 & 0 & \frac{1}{4}%
\end{bmatrix}%
\begin{bmatrix}
1 & 0 & 0 & 0\\
0 & 0 & 0 & 0\\
0 & 0 & 1 & 0\\
0 & 0 & 0 & 0
\end{bmatrix}
=\allowbreak%
\begin{bmatrix}
\frac{1}{4} & 0 & \frac{1}{4} & 0\\
0 & 0 & 0 & 0\\
\frac{1}{4} & 0 & \frac{1}{4} & 0\\
0 & 0 & 0 & 0
\end{bmatrix}
$

$%
\begin{bmatrix}
0 & 0 & 0 & 0\\
0 & 1 & 0 & 0\\
0 & 0 & 0 & 0\\
0 & 0 & 0 & 1
\end{bmatrix}%
\begin{bmatrix}
\frac{1}{4} & \frac{1}{4} & \frac{1}{4} & 0\\
\frac{1}{4} & \frac{1}{4} & \frac{1}{4} & 0\\
\frac{1}{4} & \frac{1}{4} & \frac{1}{4} & 0\\
0 & 0 & 0 & \frac{1}{4}%
\end{bmatrix}%
\begin{bmatrix}
0 & 0 & 0 & 0\\
0 & 1 & 0 & 0\\
0 & 0 & 0 & 0\\
0 & 0 & 0 & 1
\end{bmatrix}
=\allowbreak%
\begin{bmatrix}
0 & 0 & 0 & 0\\
0 & \frac{1}{4} & 0 & 0\\
0 & 0 & 0 & 0\\
0 & 0 & 0 & \frac{1}{4}%
\end{bmatrix}
$

$%
\begin{bmatrix}
\frac{1}{4} & 0 & \frac{1}{4} & 0\\
0 & 0 & 0 & 0\\
\frac{1}{4} & 0 & \frac{1}{4} & 0\\
0 & 0 & 0 & 0
\end{bmatrix}
+\allowbreak%
\begin{bmatrix}
0 & 0 & 0 & 0\\
0 & \frac{1}{4} & 0 & 0\\
0 & 0 & 0 & 0\\
0 & 0 & 0 & \frac{1}{4}%
\end{bmatrix}
=\allowbreak%
\begin{bmatrix}
\frac{1}{4} & 0 & \frac{1}{4} & 0\\
0 & \frac{1}{4} & 0 & 0\\
\frac{1}{4} & 0 & \frac{1}{4} & 0\\
0 & 0 & 0 & \frac{1}{4}%
\end{bmatrix}
$

$\hat{\rho}=\rho\left(  \left\{  ac,bd\right\}  \vee\left\{  abc,d\right\}
\right)  =\rho\left(  \left\{  ac,b,d\right\}  \right)  $.
\end{center}

The logical entropy of $\rho\left(  \left\{  abc,d\right\}  \right)  $ is

\begin{center}
$h\left(  \rho\left(  \left\{  abc,d\right\}  \right)  \right)
=1-\operatorname*{tr}\left[  \rho\left(  \left\{  abc,d\right\}  \right)
^{2}\right]  =1-\frac{10}{16}=\frac{3}{8}$
\end{center}

\noindent and the logical entropy of $\hat{\rho}=\rho\left(  \left\{
ac,b,d\right\}  \right)  $ is

\begin{center}
$h\left(  \rho\left(  \left\{  ac,b,d\right\}  \right)  \right)
=1-\operatorname*{tr}\left[  \rho\left(  \left\{  ac,b,d\right\}  \right)
^{2}\right]  =1-\frac{3}{8}=\frac{5}{8}$.
\end{center}

In the transition from $\rho\left(  \left\{  abc,d\right\}  \right)  $ to
$\hat{\rho}=\rho\left(  \left\{  ac,b,d\right\}  \right)  $, there were four
entries of $\frac{1}{4}$ that were zeroed, so the sum of their squares is:
$4\times\frac{1}{16}=\frac{1}{4}$ which by the Corollary equals the increase
in logical entropy; $\frac{5}{8}-\frac{3}{8}=\frac{1}{4}$.

Repeated joins, i.e., repeated intersections of blocks of different
partitions, may eventually reach the discrete partition $\mathbf{1}_{U}$ whose
density matrix is diagonal having no non-zero off-diagonal elements, i.e., all
possible distinctions have been made (like a completely decomposed mixed state
in QM). A set of partitions on $U$ whose join is the discrete partition is
said to be \textit{complete}, and it is the partition-logical analogue of
Dirac's complete set of commuting observables (CSCO) \cite{dirac:principles}.

\section{The Yoga of Linearization: From sets to vector spaces}

Our thesis is that the mathematics of QM is essentially the mathematics of
partitions linearized to (Hilbert) vector spaces. There is a semi-algorithmic
method--part of the folklore of mathematics (see Weyl's use of it below) to
linearize concepts using sets (e.g., partitions) to the corresponding concepts
over vector spaces--a method that Gian-Carlo Rota might call a ``yoga" \cite[p.
251]{rota:indiscrete}. The idea is based on taking the vector space concept
corresponding to the notion of a set as a basis set of the space. Then the
yoga is:

\begin{center}
For any given set-concept, apply it to a basis set

and whatever is linearly generated is the corresponding vector space concept.

\textit{The Yoga of Linearization}.
\end{center}

In applying the Yoga, we take $U$ as being first a set and then a basis set of
a vector space over a field $\Bbbk$.\footnote{There is the linearization map
(functor) which takes a set $U$ to the vector space $%
\mathbb{C}
^{U}$ where $u$ lifts to the basis vector $\delta_{u}=\chi_{\left\{
u\right\}  }:U\rightarrow%
\mathbb{C}
$, but we apply the Yoga to many different concepts.} For instance, the set
concept of a subset $S$ when applied to a basis set generates a subspace
$\left[  S\right]  $. The cardinality of the $U$ gives the dimension of the
space $\left[  U\right]  $ generated by the basis set $U$ as shown in Table 4.

\begin{center}%
\begin{tabular}
[c]{|c|c|}\hline
Set concept & Vector-space concept\\\hline\hline
Universe set $U$ & $U$ basis set of a space $\left[  U\right]  =V$ over
$\Bbbk$\\\hline
Cardinality of the set $U$ & Dimension of the space $V$\\\hline
Subset $S$ of the set $U$ & Subspace $\left[  S\right]  $ of the space
$V$\\\hline
\end{tabular}

Table 4: Some initial applications of the Yoga.
\end{center}

\noindent In particular, a singleton subset (representing an eigenstate as in
Figure 4) generates a one-dimensional ray. The previous set example of joining
the definite states $\left\{  a\right\}  $ and $\left\{  c\right\}  $ to form
the indefinite superposition $\left\{  a,c\right\}  $ would linearize to the
superpositions that we might represent as $k\left\vert a\right\rangle
+k^{\prime}\left\vert c\right\rangle $ for $k,k^{\prime}\in\Bbbk$. It should
also be noted that the treatment of a basis element as definite and a linear
combination of basis vectors as indefinite is always a description relative to
that basis. If the vector space $V$ has inner products like a Hilbert space,
then we will assume $U$ is an orthonormal basis.

The Yoga may sometimes require a choice. Consider the set concept of a
numerical attribute $f:U\rightarrow\Bbbk$ taking values in a field $\Bbbk$.
Taking $U$ as a basis set for a vector space $V$ over $\Bbbk$, it defines both
a linear functional $\hat{f}:V\rightarrow\Bbbk$ and a diagonalizable linear
operator: $F:V\rightarrow V$ generated by $Fu_{i}=f\left(  u_{i}\right)
u_{i}$ on the basis vectors $u_{i}\in U$. The numerical attribute
$f:U\rightarrow\Bbbk$ is recovered from the linear operator as the eigenvalue
function assigning eigenvalues to the eigenvector basis set $U$.

For the purposes of extending the mathematics of set partitions to vector
spaces to obtain the mathematical tools of QM, it is the \textit{operator} $F$
that is used in the Yoga, not the linear functional. The elements of the basis
set $U$ are a basis of eigenvectors and the values of the numerical attribute
are the eigenvalues of the operator $F$. Two numerical attributes defined on
the \textit{same} set $U$ would generate two linear operators that commute
with the basis set $U$ as a basis of simultaneous eigenvectors as illustrated
in Table 5.

\begin{center}%
\begin{tabular}
[c]{|c|c|}\hline
Set concept & Vector-space concept\\\hline\hline
$f:U\rightarrow\Bbbk$ & $F:V\rightarrow V$ by $Fu=f\left(  u\right)
u$\\\hline
$g:U\rightarrow\Bbbk$ & $G:V\rightarrow V$ by $Gu=g\left(  u\right)
u$\\\hline
{\small \ }$f,g${\small \ on same set }$U$ & {\small \ }$F,G$%
{\small \ commuting with basis }$U${\small \ of simult. eigenvectors}\\\hline
\end{tabular}

Table 5: Operators corresponding to numerical attributes.
\end{center}

The inverse-image of $f:U\rightarrow\Bbbk$ is a partition on $U$ where each
block $f^{-1}\left(  k\right)  $ is associated with a distinct $k\in
f(U)\subseteq\Bbbk$. What is the vector-space version of a set partition? If
we had taken the vector-space analog of $f:U\rightarrow\Bbbk$ as the linear
functional $\hat{f}:V\rightarrow\Bbbk$, then the corresponding vector space
concept would be the inverse-image of the functional which is a special type
of \textit{set} partition on $V$ called a \textit{commuting} or
\textit{permutable} partition (much studied by Gian-Carlo Rota and earlier by
Dubreil and Dubreil-Jacotin \cite{dub:cer}).

Following our choice, the vector-space notion of a partition is a
\textit{direct-sum decomposition}.\footnote{In view of the duality between
subsets and set partitions, there is the induced duality between subspaces and
direct-sum decompositions. Hence the Birkhoff-von-Neumann quantum logic of
subspaces \cite{birk-vn:qm-logic} has a dual quantum logic of DSDs
\cite{ell:dsds}.} Each block in the set partition $\left\{  f^{-1}\left(
k\right)  \right\}  _{k\in f\left(  U\right)  }$ generates a subspace which is
just the eigenspace $V_{k}$ of the linear operator $F:V\rightarrow V$
determined by the eigenvalue $k\in f\left(  U\right)  \subseteq\Bbbk$, and the
vector space $V$ is the direct sum of the eigenspaces $V=\oplus_{k\in f\left(
U\right)  }V_{k}$.

A direct-sum decomposition (DSD) $\left\{  V_{k}\right\}  _{k\in f\left(
U\right)  }$ is characterized by the fact that every nonzero vector $v\in V$
can represented in exactly one way as the sum $v=\sum_{k\in f\left(  U\right)
}v_{k}$ of nonzero vectors $v_{k}\in V_{k}$. The vectors $v_{k}$ are the
projections $P_{k}\left(  v\right)  $ of $v$ to the subspaces $V_{k}$. What
happens if we run the Yoga backwards? Does that property of DSDs also
characterize set partitions? A set partition is usually defined as a set of
nonempty subsets $\left\{  B_{1},B_{2},...,B_{m}\right\}  $ that are disjoint
and jointly exhaustive. Each block $B_{j}$ defines a projection operator
$B_{j}\cap():\wp\left(  U\right)  \rightarrow\wp\left(  U\right)  $ on the
powerset $\wp\left(  U\right)  $ as a vector space $%
\mathbb{Z}
_{2}^{n}$ over $%
\mathbb{Z}
_{2}$ (with set addition as the symmetric difference \cite{ell:qm-sets}) down
to the subspace $\wp\left(  B_{j}\right)  $. Then we have the result:

\begin{center}
$\left\{  B_{1},B_{2},...,B_{m}\right\}  $ is a set partition of $U$

if and only if

every nonempty subset $S\subseteq U$, is uniquely expressed as the union of
subsets of the $B_{j}$, $j=1,...,m$

if and only if

$\left\{  \wp\left(  B_{j}\right)  \right\}  _{j=1}^{m}$ is a DSD of
$\wp\left(  U\right)  \cong%
\mathbb{Z}
_{2}^{n}$.
\end{center}

This running of the Yoga backwards raises the question of what is the
set-analogue of the notion of an eigenvector and eigenvalue. For $k\in\Bbbk$
and $S\subseteq U$, let ``$kS$" stand for ``the value $k$ assigned to the
elements of $S$". Then we have:

\begin{center}
The eigenvector/eigenvalue equation for $f:U\rightarrow\Bbbk$:

$f\upharpoonright S=kS$ in analogy with $Fu=ku$.
\end{center}

\noindent Thus the set-notion of an eigenvector is just a constant set and the
set notion of an eigenvalue is that constant value on a constant set as
illustrated in Table 6. 

A characteristic function $\chi_{S}:U\rightarrow\left\{  0,1\right\}
\subseteq\Bbbk$ applied to a basis set for $V$ linearizes to a projection
operator $P_{\left[  S\right]  }:V\rightarrow V$ defined by $P_{\left[
S\right]  }u=\chi_{S}\left(  u\right)  u$. The constant sets of $\chi_{S}$ are
$S$ and $S^{c}=U-S$ and the eigenspaces of $P_{\left[  S\right]  }$ are
$\left[  S\right]  $ and $\left[  S^{c}\right]  $ with the respective
eigenvalues of $1$ and $0$. In general, $f:U\rightarrow\Bbbk$ linearized to
$F$ defined by $Fu=f\left(  u\right)  u$ with the eigenspaces of $F$ being
$V_{k}=\left[  f^{-1}\left(  k\right)  \right]  $ for $k\in f\left(  U\right)
$. If $P_{k}:V\rightarrow V$ is the projection operator to $V_{k}$, then the
spectral decomposition of $F$ is $\sum_{k}kP_{k}$ and the corresponding
`spectral decomposition' for $f$ is: $f=\sum_{k\in f\left(  U\right)  }%
k\chi_{f^{-1}\left(  k\right)  }:U\rightarrow\Bbbk$.

\begin{center}%
\begin{tabular}
[c]{|c|c|}\hline
Set concept & Vector-space concept\\\hline\hline
{\small Partition} $\left\{  f^{-1}\left(  k\right)  \right\}  _{k\in f\left(
U\right)  }$ & {\small DSD }$\left\{  V_{k}\right\}  _{k\in f\left(  U\right)
}${\small :} $V=\oplus_{k\in f\left(  U\right)  }V_{k}$\\\hline
$f\upharpoonright S=kS$ & $Fu_{i}=ku_{i}$\\\hline
Constant set $S$\ of $f$ & Eigenvector $u_{i}$\ of $F$\\\hline
Value $k$\ on constant set $S$ & Eigenvalue $k$\ of eigenvector $u_{i}%
$\\\hline
Characteristic fcn. $\chi_{S}:U\rightarrow\Bbbk$ & Projection op. $P_{\left[
S\right]  }:V\rightarrow V$\\\hline
Spectral decomp. $f=\sum_{k\in f\left(  U\right)  }k\chi_{f^{-1}\left(
k\right)  }$ & Spectral decomp. $F=\sum_{k}kP_{k}$\\\hline
Set of $k$-constant sets $\wp\left(  f^{-1}\left(  k\right)  \right)  $ &
Eigenspace of $k$-eigenvectors $V_{k}$\\\hline
Partition join $\pi\vee\sigma$ & L\"{u}ders op. $\sum_{k\in f\left(  U\right)
}P_{k}\rho P_{k}$\\\hline
\end{tabular}

Table 6: Corresponding concepts based on partition of a set and direct-sum
decomposition of a vector space.
\end{center}

What is the vector space version of the Cartesian or direct product of sets
$U\times U^{\prime}$? One yoga might say the direct product of vector spaces
$V\times W$. But our Yoga is apply the set concept to basis sets and see what
it generates. Let $U$ be a basis for $V$ and $U^{\prime}$ be a basis for $W$,
both over the same field, then applying the Cartesian product to the basis
sets gives $U\times U^{\prime}$ and it (bi)linearly generates the tensor
product $V\otimes W$ with the ordered pair $\left(  u,u^{\prime}\right)  $
elements of $U\times U^{\prime}$ corresponding to the basis elements $u\otimes
u^{\prime}$ of $V\otimes W$ as given in Table 7.

\begin{center}%
\begin{tabular}
[c]{|c|c|}\hline
Set concept & Vector-space concept\\\hline\hline
Cartesian product $U\times U^{\prime}$ & Tensor product $V\otimes W$\\\hline
$\left(  u,u^{\prime}\right)  $ element of $U\times U^{\prime}$ & Basis
element $u\otimes u^{\prime}$ of $V\otimes W$\\\hline
\end{tabular}

Table 7: Cartesian product of sets corresponds to the tensor product of vector spaces.
\end{center}

\section{The mathematics of quantum mechanics}

\subsection{Commuting and non-commuting observables}

One of the characteristic features of QM mathematics is the possibility that
observables (expressed as self-adjoint operators or matrices) do not
commute--which at first does not seem related to the partition math of
indefiniteness. But the vector space version of a partition on a set is a
direct-sum decomposition of a vector space. Given two self-adjoint operators
$F,G:V\rightarrow V$, let $\left\{  V_{i}\right\}  _{i\in I}$ be the DSD of
eigenspaces for $F$ and $\left\{  W_{j}\right\}  _{j\in J}$ be the DSD of
eigenspaces for $G$. As we saw for quantum measurement, the relevant partition
operation is the join, so we may mimic the join operation with the two DSDs.
This join-like operation yields the set of non-zero vector spaces $\left\{
V_{i}\cap W_{j}\right\}  $ which are the subspaces spanned by the simultaneous
eigenvectors of $F$ and $G$. The join of two partitions on the \textit{same}
set yields a partition of that set. Let $\mathcal{SE}$ be the subspace of $V$
spanned by the non-zero subspaces $\left\{  V_{i}\cap W_{j}\right\}  $, i.e.,
the subspace spanned by the simultaneous eigenvectors of $F$ and $G$. The
point is that $\mathcal{SE}$ need not be the whole space. The condition
specifying whether $F$ and $G$ commute or not is exactly the condition that
$\mathcal{SE}=V$ or not. The \textit{commutator} of $F$ and $G$ is: $\left[
F,G\right]  =FG-GF:V\rightarrow V$, and as a linear operator on $V$, the
commutator has a kernel $\ker\left[  F,G\right]  $ which is the subspace of
vectors $v$ such that $\left[  F,G\right]  v=0$.

\begin{proposition}
$\mathcal{SE}=\ker\left(  \left[  F,G\right]  \right)  $.
\end{proposition}

Proof: Let $F,G:V\rightarrow V$ be two self-adjoint operators on a finite
dimensional vector space $V$ and let $v$ be a simultaneous eigenvector of the
operators, i.e., $Fv=\lambda v$ and $Gv=\mu v$. Then $\left[  F,G\right]
\left(  v\right)  =\left(  FG-GF\right)  \left(  v\right)  =\left(  \lambda
\mu-\mu\lambda\right)  v=0$ so the space $\mathcal{SE}$ spanned by the
simultaneous eigenvectors is contained in the kernel $\ker\left(  \left[
F,G\right]  \right)  $, i.e., $\mathcal{SE}\subseteq\ker\left(  \left[
F,G\right]  \right)  $. Conversely, if we restrict the two operators to the
subspace $\ker\left(  \left[  F,G\right]  \right)  $, then the restricted
operators commute on that subspace. Then it is a standard theorem of linear
algebra \cite[p. 177]{hoffman-kunze:la-1st-ed} that the subspace $\ker\left(
\left[  F,G\right]  \right)  $ is spanned by simultaneous eigenvectors of the
two restricted operators. But if a vector is a simultaneous eigenvector for
the two operators restricted to a subspace, they are the same for the
operators on the whole space $V$, since the two conditions $Fv=\lambda v$ and
$Gv=\mu v $ only involves vectors in the subspace. Hence $\ker\left(  \left[
F,G\right]  \right)  \subseteq\mathcal{SE}$. $\square$

Since the condition that the operators commute or not is $\ker\left(  \left[
F,G\right]  \right)  =V$ or not, it is equivalent to $\mathcal{SE}=V$ or not,
so the commutativity condition on operators is captured by the mathematics of
the vector-space version of partitions, i.e., DSDs. And the further condition
of the operators being \textit{conjugate} is when $\mathcal{SE}=\mathbf{0}$
(the subspace consisting of only the zero vector). The Heisenberg
"uncertainty" principle is somewhat misnamed since ``uncertainty" may imply a
subjective uncertainty instead of objective indefiniteness. The
"indefiniteness principle" or ``indeterminacy principle" might be a better
name.\footnote{Heisenberg's German word was ``Unbestimmtheit" which could well
be translated as ``indefiniteness" or ``indeterminacy" rather than
"uncertainty."} In any case, since conjugate observables have no (non-zero)
simultaneous eigenvectors, $\mathcal{SE}=\mathbf{0}$, if a system is in an
eigenstate of one observable, it cannot be in an eigenstate of the other observable.

Since we have shown how the mathematics of indefiniteness can be translated
into vector spaces, it might be noted this conjugacy is not a peculiarly
quantum concept about operators in Hilbert spaces but can occur in quite
simple vector spaces such as $\wp\left(  U\right)  \cong%
\mathbb{Z}
_{2}^{n}$ (where set addition in $\wp\left(  U\right)  $ is symmetric
difference \cite{ell:qm-sets}). In particular, for $n$ an even number as with
$U=\left\{  a,b,c,d\right\}  $, the $U$-basis set $\left\{  \left\{
a\right\}  ,\left\{  b\right\}  ,\left\{  c\right\}  ,\left\{  d\right\}
\right\}  $ for $\wp\left(  U\right)  \cong%
\mathbb{Z}
_{2}^{4}$ has a conjugate basis of:

\begin{center}
$\left\{  \hat{a}\right\}  =\left\{  b,c,d\right\}  ,\left\{  \hat{b}\right\}
=\left\{  a,c,d\right\}  ,\left\{  \hat{c}\right\}  =\left\{  a,b,d\right\}
,\left\{  \hat{d}\right\}  =\left\{  a,b,c\right\}  $
\end{center}

\noindent which constitutes the $\hat{U}$-basis. Two numerical attributes
$f:U\rightarrow%
\mathbb{R}
$ and $g:\hat{U}\rightarrow%
\mathbb{R}
$ with distinct values on each basis set (e.g., $1,2,3,4$) would define the
two DSDs on $\wp\left(  U\right)  \cong%
\mathbb{Z}
_{2}^{4}$ which, expressed in the $U$-basis as the computational basis, are:

\begin{center}
DSD for $f$ is $\left\{  \left\{  0,\left\{  a\right\}  \right\}  ,\left\{
0,\left\{  b\right\}  \right\}  ,\left\{  0,\left\{  c\right\}  \right\}
,\left\{  0,\left\{  d\right\}  \right\}  \right\}  $

DSD for $g$ is $\left\{  \left\{  0,\left\{  b,c,d\right\}  \right\}
,\left\{  0,\left\{  a,c,d\right\}  \right\}  ,\left\{  0,\left\{
a,b,d\right\}  \right\}  ,\left\{  0,\left\{  a,b,c\right\}  \right\}
\right\}  $.
\end{center}

\noindent Clearly, there no non-zero eigenvectors in common between the two
DSDs so the numerical attributes are conjugate. If the state was in an
eigenstate of one attribute such as $\left\{  a\right\}  $, then it is in an
indefinite superposition $\left\{  \hat{b}\right\}  +\left\{  \hat{c}\right\}
+\left\{  \hat{d}\right\}  =\left\{  a\right\}  $ according to the conjugate
basis (and vice-versa). In Figure 4, we illustrated the set-level pure and
mixed states using the lattice $\Pi\left(  U\right)  $. The conjugate $\hat
{U}$-basis also has a similar lattice $\Pi\left(  \hat{U}\right)  $ and moving
to a more definite state $\left\{  a\right\}  +\left\{  b\right\}  +\left\{
c\right\}  =\left\{  a,b,c\right\}  \rightsquigarrow\left\{  a\right\}  $ in
the $U$-basis would correspond to moving to a less definite state in the
conjugate basis $\hat{U}$, e.g., $\left\{  a,b,c\right\}  =\left\{  \hat
{d}\right\}  \rightsquigarrow\left\{  \hat{b},\hat{c},\hat{d}\right\}
=\left\{  a\right\}  $, as one would expect for conjugate bases.

It might be noted that these numerical attributes cannot be repackaged as
linear operators with the eigenvalues in the base field since the only such
linear operators on $%
\mathbb{Z}
_{2}^{n}$ are projection operators with eigenvalues $0$ or $1$. But all the
concepts of compatibility, i.e., $\mathcal{SE}=V$, incompatibility, i.e.,
$\mathcal{SE}\neq V$, and conjugacy, i.e., $\mathcal{SE}=\mathbf{0}$, can be
defined using the vector-space partitions, i.e., DSDs, in $%
\mathbb{Z}
_{2}^{4}$. Thus the mathematics behind ``non-commutativity" in QM is not about
operators \textit{per se}, but about the underlying vector space partitions or
DSDs--as was to be shown.

Commuting observables are like ordinary numerical attributes on the
\textit{same} set $U$--in that case a basis of simultaneous eigenvectors. It
is only when $\mathcal{SE}=V$ that the join-like operation taking non-zero
intersections of the eigenspaces can properly be called the \textit{join} of
the DSDs, otherwise it is a join-like (or proto-join)\textit{ }operation when
$\mathcal{SE}\neq V$. As Hermann Weyl put it: ``Thus combination [join] of two
gratings [vector space partitions] presupposes commutability...". \cite[p.
257]{weyl:phil}

A set of ordinary set partitions on the same universe $U$ is said to be
\textit{complete} if their join is the discrete partition $\mathbf{1}%
_{U}=\left\{  \left\{  u\right\}  \right\}  _{u\in U}$ where the subsets in
the join have cardinality one. Numerical attributes defined on the same set
are \textit{compatible}, and if they defined a complete set of partitions,
they would be a complete set of compatible attributes (CSCA). If the
partitions arose as the inverse-images of numerical attributes (or random
variables), then each element in $U$ would be characterized by the ordered set
of values of the attributes.

Similarly a set of commuting operators is said be \textit{complete }(a CSCO)
\cite{dirac:principles} if all the non-zero intersections of all their
eigenspaces, i.e., the subspaces in the join, are of dimension one. Then each
of the simultaneous eigenvectors is uniquely characterized by the ordered set
of eigenvalues of those intersected eigenspaces. These results are summarized
in Table 8.

\begin{center}%
\begin{tabular}
[c]{|c|c|}\hline
Set concept & Vector-space concept\\\hline\hline
Set partition & Direct-sum decomposition\\\hline
Partitions on $U\neq U^{\prime}$, $\left\vert U\right\vert =\left\vert
U^{\prime}\right\vert $ & DSDs on $V$ with $\mathcal{SE}\neq V$\\\hline
Partitions on same set $U$ & DSDs on $V$ with $\mathcal{SE}=V$\\\hline
Join of partitions on same set $U$ & Join of DSDs is DSD of $V$\\\hline
CSCA & CSCO\\\hline
\end{tabular}

Table 8: Set and vector-space versions of commutativity.
\end{center}

\subsection{The two types of quantum processes}

Quantum concepts need to be `seen' in a certain way to see the underlying
mathematics of indefiniteness and definiteness as shown in the case of
non-commuting operators. At first glance, the Schr\"{o}dinger equation to
describe the evolution of an isolated system seems to have nothing to do with
distinctions. Von Neumann classified quantum processes into Type 1
(measurement) and Type 2 (evolution described by the Schr\"{o}dinger
equation). We have seen that a measurement or Type 1 process creates
distinctions so the natural characterization of the Type 2 processes would be
ones that make no distinctions.

The extent to which two quantum states are indistinct or distinct is given by
their inner product, i.e., their overlap. When their inner product is zero,
then there is zero indistinctness or zero overlap between the states, i.e.,
they are fully distinct. Hence the natural characterization of the Type 2
processes as not changing the indistinctness or distinctness between quantum
states would a process that preserves inner products, i.e., a unitary
transformation. Hence the division of quantum processes into Type 1 and Type 2
is just the division between the processes that makes distinctions and those
that don't.

What about the Schr\"{o}dinger equation? The connection between unitary
transformations and the solutions to the Schr\"{o}dinger equation is given by
Stone's Theorem \cite{stone:thm}: there is a one-to-one correspondence between
strongly continuous $1$-parameter \textit{unitary} groups $\left\{
U_{t}\right\}  _{t\in%
\mathbb{R}
}$ and self-adjoint operators $H$ (Hamiltonian) on the Hilbert space so that
$U_{t}=e^{iHt}$ (solutions of the Schr\"{o}dinger equation).

\subsection{Measurement and the collapse postulate}

We have seen that quantum measurements create distinctions. Richard Feynman
was perhaps the quantum theorist who most emphasized measurement as making
distinctions. When a superposition of eigenstates undergoes an interaction, is
there a distinction made in principle between the superposed eigenstates in
the interaction? If the eigenstates are distinguished by the interaction, then
a measurement takes place, the superposition is reduced (i.e., the so-called
"wave function" collapses), and the probability of a later final state will
add the probabilities (rather than amplitudes) of the eigenstates leading to
the outcome. If there is no differences or distinctions between the superposed
eigenstates undergoing the interaction, then no measurement takes place and
the amplitudes are added.

\begin{quotation}
\noindent If you could, \textit{in principle}, distinguish the alternative
\textit{final} states (even though you do not bother to do so), the total,
final probability is obtained by calculating the \textit{probability} for each
state (not the amplitude) and then adding them together. If you
\textit{cannot} distinguish the final states \textit{even in principle}, then
the probability amplitudes must be summed before taking the absolute square to
find the actual probability.\cite[p. 3-9]{feynman:vIII}
\end{quotation}

Feynman thus answers a question posed in the literature where the key concepts
of distinguishability and indistinguishability are not used.

\begin{quotation}
It indeed seems necessary to admit that \textquotedblleft
measurements\textquotedblright\ are ubiquitous, and occur even in places and
times where there are no human experimenters. But it also seems hopeless to
think that we will be able to give an appropriately sharp answer to the
question of what, exactly, differentiates the `ordinary' processes (where the
usual dynamical rules apply) from the `measurement-like' processes (where the
rules momentarily change). \cite[p. 64]{norsen:fqm}

\noindent\ [I]t seems unbelievable that there is a fundamental distinction
between \textquotedblleft measurement\textquotedblright\ and \textquotedblleft
non-measurement\textquotedblright\ processes. Somehow, the true fundamental
theory should treat all processes in a consistent, uniform fashion. \cite[p.
245]{norsen:fqm}
\end{quotation}

\noindent The ``fundamental distinction" is between processes where, in the
interaction, distinctions are made or are not made between the eigenstates in
the superposition.

Feynman gives an example of a measurement entirely at the quantum level, and
thus he undercuts the long and tortured discussion about measurement as
involving a macroscopic apparatus. When a particle scatters off the atoms in a
crystal, the question of whether or not it should be treated as a
superposition of scattering off the different atoms or as a mixture of
scattering off of particular atoms with certain probabilities--hinges on
distinguishability. If there was no distinction between scattering off
different atoms, then no `measurement' took place in the interaction and the
superposition pure state evolves as a pure state. But if there was some
distinction caused by scattering off an atom, then the result is the mixed
state of scattering off the different atoms with different probabilities. For
instance, if all the atoms had spin down and scattering off an atom flipped
the spin, then a distinction was made so that constituted a measurement. It
should be noted that this and other examples of Feynman \cite{feynman:vIII}
involve only quantum level interactions and thus have nothing to do with the
"shifty split" \cite{bell:againstm} between microscopic and macroscopic, and
thus are independent of the notion of ``decoherence" based on interactions with
macroscopic systems (e.g., \cite{zurek:decoh}).

For instance in the double-slit experiment, if no distinction is made between
the particle going through one slit or the other (i.e., no detectors at the
slits), then the two parts in the superposition schematically represented by:

\begin{center}
$\left\vert \text{going through slit }1\right\rangle +\left\vert \text{going
through slit }2\right\rangle $
\end{center}

\noindent evolve unitarily and will show interference effects. The mathematics
of the unitary (i.e., no distinctions) evolution over the complex numbers will
have a complex-valued ``wave interpretation" (Stone's Theorem) without there
being any physical waves; the interference results from the addition of
vectors representing the parts of the evolving superposition.\footnote{This is
particularly clear in the pedagogical model of the double-slit experiment in
QM over $%
\mathbb{Z}
_{2}$ \cite{ell:qm-sets} where the interference in the evolving $0,1$-vectors
has no resemblence to waves.}

There being no distinctions involved, the evolving indefinite superposition
does not reach the level of definiteness of the particle going through one
slit or the other--which conflicts with our `classical' intuitive
always-definite view of the particle's trajectory. A better but still crude
intuitive picture would be to recognize the \textit{levels of indefiniteness}
at the set level in the lattice of partitions (e.g., Figure 4)--which would
prefigure levels of indefiniteness at the quantum level. There is the
completely indefinite state represented by the indiscrete partition at the
bottom of the lattice, and then moving up to partitions with each block
representing an equally or more definite superposition, and finally reaching
the discrete partition at the top where each block is a singleton `eigenstate'
analogous to the mixed state of fully decomposed eigenstates represented by a
diagonal density matrix.

The difference between an interaction that constitutes a measurement or not is
whether or not any distinction is made between the different superposed
eigenstates undergoing the interaction. Hence Feynman's implicit rule about
state reduction might be paraphrased:

\begin{center}
\textit{If the interaction would make distinctions, then distinctions are
made.}
\end{center}

\noindent In other words, if the interaction makes a difference between the
superposed eigenstates, then the superposition decoheres with a
(indefinite-to-definite) state reduction to one of those eigenstates.

One image for the measurement process is a `shapeless' or indefinite blob of
dough which then passes through a sieve or grating and acquires a definite
polygonal shape as illustrated in Figure 7.%

\begin{figure}[h!]
	\centering
	\includegraphics[width=0.5\linewidth]{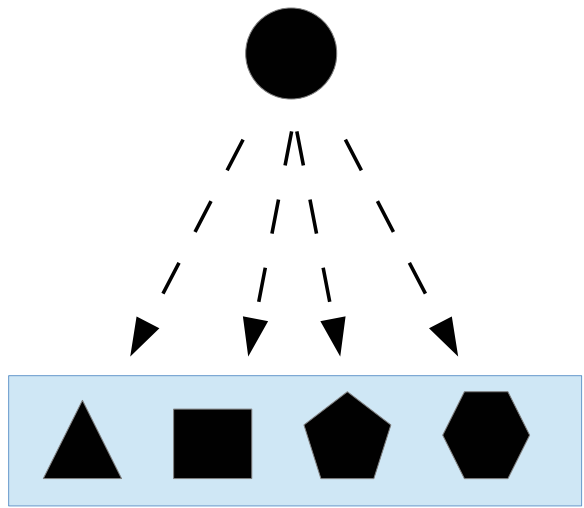}
	\caption{Measurement as an `indefinite' shape passing through a sieve to get a definite polygonal shape}
	\label{fig:fig7doughsieve}
\end{figure}

\noindent This sort of grating or sieve imagery has been used before. In his
popular writing, Arthur Eddington used the sieve metaphor:

\begin{quotation}
\noindent In Einstein's theory of relativity the observer is a man who sets
out in quest of truth armed with a measuring-rod. In quantum theory he sets
out armed with a sieve.\cite[p. 267]{edd:pathways}
\end{quotation}

Hermann Weyl cited that passage \cite[p. 255]{weyl:phil} in his expositional
concept of gratings. Then Weyl, in effect, used the Yoga of Linearization by
taking \textit{both} set partitions and vector space partitions (direct-sum
decompositions) as the respective types of gratings.\cite[pp. 255-257]%
{weyl:phil} He started with a numerical attribute on a set (or ``aggregate" in
older terminology), e.g., $f:U\rightarrow%
\mathbb{R}
$, which defined the set partition or ``grating" \cite[p. 255]{weyl:phil} with
blocks having the same attribute-value, e.g., $\left\{  f^{-1}\left(
r\right)  \right\}  _{r\in f\left(  U\right)  }$. Then he implicitly applied
the Yoga to reach the QM case where the universe set, e.g., $U=\left\{
u_{1},...,u_{n}\right\}  $, or ``aggregate of $n$ states has to be replaced by
an $n$-dimensional Euclidean vector space" \cite[p. 256]{weyl:phil}. Then the
notion of a vector space partition or ``grating" in QM is a ``splitting of the
total vector space into mutually orthogonal subspaces" so that ``each vector
$\overrightarrow{x}$ splits into $r$ component vectors lying in the several
subspaces" \cite[p. 256]{weyl:phil}, i.e., a direct-sum decomposition of the
space. After thus referring to a partition and a DSD as a ``grating" or
"sieve," Weyl notes that ``Measurement means application of a sieve or grating"
\cite[p. 259]{weyl:phil}, i.e., an interaction that makes distinctions and
thus forces an indefinite-to-definite transition, e.g., Figure 7.

Our overall goal is to show that the mathematics of QM is the
mathematics in distinctions and indistinctions or definiteness and
indefiniteness expressed in terms of vector spaces. The vision of realism
based on objective indefiniteness is juxtaposed to our ordinary intuitive idea
that reality is definite-all-the-way-down. One aspect of ``the measurement
problem" has been the lack of any mathematical description of how a quantum
system goes from an objectively indefinite superposition state to a more
definite state during a measurement. But that question seems to arise out of
imposing the\ fully definite framework as if it was only a transition from a
perfectly definite but rather featureless state to another definite state with
more discernible features--like classically going from a blank sheet of paper
to a sheet with figures on it. But if quantum reality consists of objectively
indefinite states, why should we expect that sort of definite-to-definite
transition between states of indefiniteness--as opposed to the notion of a
genuine quantum jump or leap? Leibniz's view of reality as being definite all
the way down (expressed in his identity of indiscernible) was also expressed
by the slogan ``\textit{Natura non facit saltus}" (Nature does not make jumps")
\cite[Bk. IV, chap. xvi]{leibniz:ne}. Hence it should not be too surprising to
find the opposite phenomena of jumps in a reality that is not always definite.
Thus the unanswered question of the mathematical description of the
`trajectory of a quantum jump' may arise from an implicit assumption that
reality is definite all the way down as in classical physics.

\subsection{Indistinguishability of particles}

The classical notion of distinguishability of particles in effect treats
partitions (or DSDs) as always being refineable or definite all the way down,
e.g., distinguishing particles by `painting different colors.'

\begin{quotation}
\noindent In quantum mechanics, however, identical particles are truly
indistinguishable. This is because we cannot specify more than a complete set
of commuting observables for each of the particles; in particular, we cannot
label the particle by coloring it blue. \cite[p. 446]{sakuri:mqm}
\end{quotation}

\noindent Hence quantum indistinguishability \textit{immediately} points to
objective indefiniteness, as opposed to ``definiteness all the way down." If
definiteness does not go ``all the way down," then the making of distinctions
has to stop at some point at which the remaining indefiniteness has to be objective.

If quantum reality is not definite-all-the-way-down, then at the level where
further definiteness stops (as it were), there are two possibilities. A
complete state description (e.g., CSCO-defined) is sufficient to limit at most
a single particle to that state or the complete description is still
insufficient to limit the number of particles in that state--in neither case
distinguishing between other particles of the same type.

This difference can be illustrated by the metaphor of different levels of
definiteness in a mailing address. In a neighborhood of only single-family
houses or vacant lots, then an address that is definite down to the street
number would be sufficient to limit one or no families to each address. But in
a neighborhood which had apartment houses, then addresses limited to the
street number in definiteness (i.e., no apartment numbers) would allow many
families at the same address.

The mathematics of quantum statistics for the two types of particles can be
developed using the standard combinatorics of balls-in-boxes--which, unlike
the usual treatment using symmetric and anti-symmetric wave functions, brings
out the underlying role of distinctions and indistinctions. There are $k$
balls (or particles) and they are indistinguishable. There are $n$ boxes (or
states) and they are distinguishable. If the complete state description (or
address) is sufficient to limit one or no balls to that state or box, then as
each ball finds an empty box, then that box is removed as a possibility for
the next ball so the number of equiprobable placements of balls in boxes is
given by the \textit{falling factorial} $n\left(  n-1\right)  ...(n-k+1)$ ($k$
terms) as pictured in Figure 8.%

\begin{figure}[h!]
	\centering
	\includegraphics[width=0.4\linewidth]{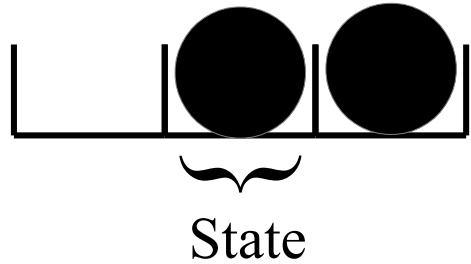}
	\caption{Fermions fill a state}
	\label{fig:fig8fermions}
\end{figure}

If, however, there can be many balls in a box, then the placement of each ball
creates two additional `slots,' before and after the ball and the number of
equiprobable slots increase with each placement to give the \textit{rising
factorial} $n\left(  n+1\right)  ...(n+k-1)$ ($k$ terms) as pictured in Figure 9.%

\begin{figure}[h!]
	\centering
	\includegraphics[width=0.5\linewidth]{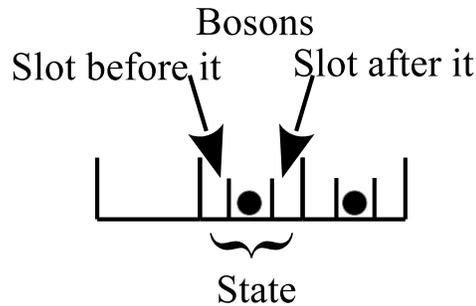}
	\caption{Each placement of a boson creates an additional possible slot}
	\label{fig:fig9bosons}
\end{figure}

\noindent In each case, since the balls are indistinguishable, the number of
equal possibilities must be divided by $k!$. Hence in the $k$-fold tensor
product, the number of equiprobable placements is the dimension of the
subspace of possible states.

\begin{itemize}
\item The dimension of the fermionic subspace the usual binomial coefficient:
\end{itemize}

\begin{center}
$\binom{n}{k}=\frac{n\left(  n-1\right)  ...(n-k+1)}{k!}=\frac{n\left(
n-1\right)  ...(n-k+1)\left(  n-k\right)  !}{k!\left(  n-k\right)  !}%
=\frac{n!}{k!\left(  n-k\right)  !}$.
\end{center}

\noindent The Fermi-Dirac statistics counts each state as having equal
probability: $1/\binom{n}{k}$.

\begin{itemize}
\item The dimension of the bosonic subspace is:
\end{itemize}

\begin{center}
$%
\genfrac{\langle}{\rangle}{0pt}{}{n}{k}%
=\frac{n\left(  n+1\right)  ...(n+k-1)}{k!}$.
\end{center}

\noindent The Bose-Einstein statistics counts each state as having equal
probability $1/%
\genfrac{\langle}{\rangle}{0pt}{}{n}{k}%
$. \cite[p. 40]{feller:vol1}

An equivalent way to enumerate the number of possible states is to enumerate
the number of functions of a certain type from balls to boxes.

\begin{itemize}
\item Fermi-Dirac statistics is based on the number of ways indistinguishable
balls (particles) are allocated to distinguishable boxes (states) using
\textit{distinction-preserving} (i.e., one-to-one) functions (so two
numerically distinct balls have to go to distinct boxes), while:

\item Bose-Einstein statistics is based on the number of ways
indistinguishable balls (particles) are allocated to distinguishable boxes
(states) using \textit{arbitrary} functions.
\end{itemize}

The classical case of Maxwell-Boltzmann (MB) statistics is where the $k$ balls
(particles) are distinguishable, the $n$ boxes (states) are distinguishable,
and the distributions of balls to boxes are by arbitrary functions. There are
$k!$ different linear orders (or permutations) of the $k$ distinguishable
particles but they are grouped into $n$ boxes with the occupation numbers of
$\theta_{1},...,\theta_{n}$ for the $n$ boxes. How many distributions are
there with those occupation numbers? The answer is still $k!$ independent of
the occupation numbers. The proof is illustrated in Figure 10 since the $n-1$
`walls' or state-dividers to make the boxes can be put in arbitrarily to get
the given occupation numbers $\theta_{1}+...+\theta_{n}=k$.%

\begin{figure}[h!]
	\centering
	\includegraphics[width=0.7\linewidth]{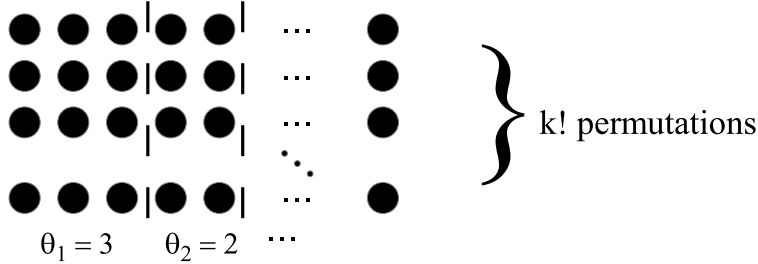}
	\caption{The number of ways $k$ balls into $n$ boxes with given occupation numbers is $k!$}
	\label{fig:fig10multinomialpix}
\end{figure}

The ordering of the balls within each box does not matter so we need to divide
through by the $\theta_{i}!$ for $i=1,...,n$ to get the total number of
possible states with those occupation numbers for the distinguishable boxes.
This gives the well-known \textit{multinomial coefficient}:

\begin{center}
$\binom{k}{\theta_{1},...,\theta_{n}}=\frac{k!}{\theta_{1}!...\theta_{n}!}$.
\end{center}

\noindent There are $n^{k}$ arbitrary functions distributing the balls in the
boxes and each distribution is classically considered equiprobable so the
probability of the given set of occupation numbers in the MB statistics is:

\begin{center}
$\binom{k}{\theta_{1},...,\theta_{n}}/n^{k}=\frac{k!}{\theta_{1}!...\theta
_{n}!}/n^{k}$.
\end{center}

The difference between MB, BE, and FD statistics can be illustrated by
computing the probability of flipping two coins of the same type. Hence there
are $k=2$ particles of the same type and $n=2$ states $\left\{  h,t\right\}  $
like two coins with heads and tails as the states. What is the probability
that one ``coin" will be ``heads" and the other ``tails"?

\begin{itemize}
\item Classical coins: $\Pr_{MB}\left(  \left\{  \left(  h,t\right)  ,\left(
t,h\right)  \right\}  \right)  =\frac{\binom{k}{\theta_{1},...,\theta_{n}}%
}{n^{k}}=\frac{\frac{k!}{\theta_{1}!...\theta_{n}!}}{n^{k}}=\frac{2}{4}%
=\frac{1}{2}$.

\item Boson coins: $\Pr_{BE}\left(  \left\{  \left(  h,t\right)  ,\left(
t,h\right)  \right\}  \right)  =\frac{k!}{n\left(  n+1\right)  ...(n+k-1)}%
=\frac{2}{2\left(  3\right)  }=\frac{1}{3}$.

\item Fermion coins: $\Pr_{FD}\left(  \left\{  \left(  h,t\right)  ,\left(
t,h\right)  \right\}  \right)  =\frac{k!}{n\left(  n-1\right)  ...(n-k+1)}%
=\frac{2}{2\left(  1\right)  }=1$.
\end{itemize}

The Pauli exclusion principle is illustrated by the fact that two fermion
coins \textit{have} to be in different states (i.e., with probability $1$). In
the case of bosons, the two classical outcomes $\left(  h,t\right)  $ and
$\left(  t,h\right)  $ differ only by a permutation of particles of the same
type so that counts only as one state out of three equiprobable states. Since
the probability of getting the same outcomes $\left(  h,h\right)  $ or
$\left(  t,t\right)  $ is $\frac{2}{3}$ in the bosonic case in comparison with
the classical MB probability of $\frac{1}{2}$, that illustrates the tendency
of bosons to ``be more social" (e.g., in terms of our metaphor, live in an
apartment house).

\subsection{Measurement with quantum logical entropy}

We can now use the Yoga to extend the `classical' notion of logical entropy
for set partitions to quantum logical entropy. The set-notion of logical
entropy was defined on set partitions determined by numerical attributes
$f:U\rightarrow\Bbbk$, and the corresponding quantum notion of logical entropy
is given by DSDs determined by self-adjoint operators $F:V\rightarrow V$ on a
Hilbert space $V$ which have a basis $U$ of eigenvectors with an eigenvalue
function $f:U\rightarrow\Bbbk$ where $Fu_{i}=f\left(  u_{i}\right)  u_{i}$. An
ordered pair of set elements $\left(  u_{i},u_{k}\right)  $ in the direct
product $U\times U$ is a distinction or dit if they have different numerical
attributes $f\left(  u_{i}\right)  \neq f\left(  u_{k}\right)  $, and,
similarly, an ordered pair of eigenvectors written $u_{i}\otimes u_{k}$ in the
tensor product $V\otimes V$ is a \textit{qudit} if they have different
eigenvalues $f\left(  u_{i}\right)  \neq f\left(  u_{k}\right)  $. Both the
set version and vector space versions of logical entropy satisfy Andrei
Kolmogorov's dictum:

\begin{quotation}
	\noindent Information theory must precede probability theory, and not be based
	on it. By the very essence of this discipline, the foundations of information
	theory have a finite combinatorial character. \cite[p. 39]{kolmogor:combfound}
\end{quotation}

\noindent In the set version, logical entropy is represented by the finite
combinatorial ditset $\operatorname*{dit}\left(  \pi\right)  \subseteq U\times
U$ and in the vector space version, the quantum logical entropy is represented
by the finite dimensional subspace $\left[  \operatorname*{qudit}(F)\right]
\subseteq V\otimes V$ (the subspace generated by the set
$\operatorname*{qudit}(F)$ of qudits $u_{i}\otimes u_{k}$ of $F$) prior to the introduction of a density matrix providing the probabilities. Neither the Shannon entropy nor the von Neumann entropy satisfy this Kolmogorov criterion. The
following Table 9 gives the set to vector space correspondence before
probabilities are introduced.

\begin{center}%
	\begin{tabular}
		[c]{|c|c|}\hline
		Classical Logical Information & Quantum Logical Information\\\hline\hline
		$f,g:U\rightarrow%
		\mathbb{R}
		$ & Commuting self-adjoint ops. $F,G$\\\hline
		$U=\left\{  u_{1},...,u_{n}\right\}  $ & ON basis simultaneous eigenvectors
		$F,G$\\\hline
		Values $\left\{  \phi_{i}\right\}  _{i=1}^{m}$ of $f$ & Eigenvalues $\left\{
		\phi_{i}\right\}  _{i=1}^{m}$ of $F$\\\hline
		Values $\left\{  \gamma_{j}\right\}  _{j=1}^{m^{\prime}}$ of $g$ & Eigenvalues
		$\left\{  \gamma_{j}\right\}  _{j=1}^{m^{\prime}}$ of $G$\\\hline
		Partition $\left\{  f^{-1}\left(  \phi_{i}\right)  \right\}  _{i=1}^{m}$ &
		Eigenspace DSD of $F$\\\hline
		Partition $\left\{  g^{-1}\left(  \gamma_{j}\right)  \right\}  _{j=1}%
		^{m^{\prime}}$ & Eigenspace DSD of $G$\\\hline
		dits of $\pi:\left(  u_{k},u_{k^{\prime}}\right)  \in U^{2}$, $f\left(
		u_{k}\right)  \neq f\left(  u_{k^{\prime}}\right)  $ & Qudits of $F$:
		$u_{k}\otimes u_{k^{\prime}}\in V\otimes V$, $f\left(  u_{k}\right)  \neq
		f\left(  u_{k^{\prime}}\right)  $\\\hline
		dits of $\sigma:\left(  u_{k},u_{k^{\prime}}\right)  \in U^{2}$, $g\left(
		u_{k}\right)  \neq g\left(  u_{k^{\prime}}\right)  $ & Qudits of $G$:
		$u_{k}\otimes u_{k^{\prime}}\in V\otimes V$, $g\left(  u_{k}\right)  \neq
		g\left(  u_{k^{\prime}}\right)  $\\\hline
		$\operatorname{dit}\left(  \pi\right)  \subseteq U\times U$ & $\left[
		\operatorname*{qudit}\left(  F\right)  \right]  $ = subspace gen. by qudits of
		$F$\\\hline
		$\operatorname{dit}\left(  \sigma\right)  \subseteq U\times U$ & $\left[
		\operatorname*{qudit}\left(  G\right)  \right]  $ = subspace gen. by qudits of
		$G$\\\hline
		$\operatorname{dit}\left(  \pi\right)  \cup\operatorname{dit}\left(
		\sigma\right)  \subseteq U\times U$ & $\left[  \operatorname*{qudit}\left(
		F\right)  \cup\operatorname*{qudit}\left(  G\right)  \right]  \subseteq
		V\otimes V$\\\hline
		$\operatorname{dit}\left(  \pi\right)  -\operatorname{dit}\left(
		\sigma\right)  \subseteq U\times U$ & $\left[  \operatorname*{qudit}\left(
		F\right)  -\operatorname*{qudit}\left(  G\right)  \right]  \subseteq V\otimes
		V$\\\hline
		$\operatorname{dit}\left(  \pi\right)  \cap\operatorname{dit}\left(
		\sigma\right)  \subseteq U\times U$ & $\left[  \operatorname*{qudit}\left(
		F\right)  \cap\operatorname*{qudit}t\left(  G\right)  \right]  \subseteq
		V\otimes V$\\\hline
	\end{tabular}

	Table 9: Ditsets and qudit subspaces without probabilities.
\end{center}

In the set case of logical entropy, we assumed a point probability
distribution $p$ on $U$ and then applied the product distribution $p\times p$
to the ditset $\operatorname*{dit}\left(  f^{-1}\right)  \subseteq U\times U$
of the inverse-image partition $\left\{  f^{-1}\left(  \phi_{i}\right)
\right\}  _{i=1}^{m}$ to get the logical entropy $h\left(  f^{-1}\right)
=p\times p\left(  \operatorname*{dit}\left(  f^{-1}\right)  \right)  $ which
was interpreted as the probability of getting different $f$-values in two
independent samples of the random variable $f$.

In the quantum case, the probabilities only enter by considering a certain
state $\psi$ with a density matrix $\rho\left(  \psi\right)  $ (represented in
the basis of $F$ eigenvectors) and then the density matrix $\rho\left(
\psi\right)  \otimes\rho\left(  \psi\right)  $ on $V\otimes V$. The
probability of getting distinct eigenvalues in two independent measurements of
the state $\psi$ by the observable $F$ is the trace of the projection
$P_{\left[  \operatorname*{qudit}\left(  F\right)  \right]  }$ to the qudit
space times the density matrix $\rho\left(  \psi\right)  \otimes\rho\left(
\psi\right)  $, which is an $n^{2}\times n^{2}$ matrix with the diagonal
entries $\left(  \rho\left(  \pi\right)  \otimes\rho\left(  \pi\right)
\right)  _{\left(  j,k\right)  ,\left(  j,k\right)  }=\rho\left(  \pi\right)
_{jj}\rho\left(  \pi\right)  _{kk}=p_{j}p_{k}$. The quantum logical entropy is then:

\begin{center}
	$h\left(  F:\psi\right)  =\operatorname{tr}\left[  P_{\left[  qudit\left(
		F\right)  \right]  }\rho\left(  \psi\right)  \otimes\rho\left(  \psi\right)
	\right]  $.
\end{center}

At the set level, we saw there were one-draw-probability versus
two-draw-probability interpretations of logical probability and logical
entropy respectively. Thus just as the quantum probability $\Pr\left(
\phi\right)  =\operatorname*{tr}\left[  P_{\phi}\rho\left(  \psi\right)
\right]  $ (where $P_{\phi}$ = projection op. to the eigenspace $V_{\phi}$ of
the eigenvalue $\phi$ and $\rho\left(  \psi\right)  $ = density matrix of
state $\psi$ represented in measurement basis) is the one-measurement
probability of getting the eigenvalue $\phi$, so the quantum logical entropy
$\operatorname{tr}\left[  P_{\left[  qudit\left(  F\right)  \right]  }%
\rho\left(  \psi\right)  \otimes\rho\left(  \psi\right)  \right]  $ is the
two-measurement probability of getting different eigenvalues in two
independent measurements of the same state.

Thus the quantum logical entropy $h\left(  F:\psi\right)  $ is defined in
terms of a state $\psi$ and measurements by the observable $F$. But in our
previous treatment of density matrices, we saw that the previous logical
entropy of set partitions could also be defined as $h\left(  \rho\left(
\pi\right)  \right)  =1-\operatorname*{tr}\left[  \rho\left(  \pi\right)
^{2}\right]  =h\left(  \pi\right)  $. Hence there is also a quantum notion
defined for any density matrix as: $h\left(  \rho\right)
=1-\operatorname*{tr}\left[  \rho^{2}\right]  $ \cite{ell:4open}. We saw that
the indiscrete partition $\mathbf{0}_{U}$ made no distinctions so $h\left(
\mathbf{0}_{U}\right)  =0$ and similarly for a pure state $\rho^{2}=\rho$ so
$h\left(  \rho\right)  =1-\operatorname*{tr}\left[  \rho^{2}\right]
=1-\operatorname*{tr}\left[  \rho\right]  =0$ since all density matrices have
trace $1$.

This correspondence is illustrated in Table 10 where $F$ and $G$ are commuting
observables so they have a basis of simultaneous eigenvectors (which is the
analogue of two numerical attributes $f,g:U\rightarrow\Bbbk$ defined on the
same set $U$).

\begin{center}%
	\begin{tabular}
		[c]{|c|c|}\hline
		`Classical' Logical Entropy & Quantum Logical Entropy\\\hline\hline
		Pure state density matrix, e.g., $\rho\left(  \mathbf{0}_{U}\right)  $ & Pure
		state density matrix $\rho\left(  \psi\right)  $\\\hline
		$U=\left\{  u_{1},...,u_{n}\right\}  $ & ON basis simultaneous eigenvectors
		$F,G$\\\hline
		$p\times p$ on $U\times U$ & $\rho\left(  \psi\right)  \otimes\rho\left(
		\psi\right)  $ on $V\otimes V$\\\hline
		$h\left(  \mathbf{0}_{U}\right)  =1-\operatorname{tr}\left[  \rho\left(
		\mathbf{0}_{U}\right)  ^{2}\right]  =0$ & $h\left(  \rho\left(  \psi\right)
		\right)  =1-\operatorname{tr}\left[  \rho\left(  \psi\right)  ^{2}\right]
		=0$\\\hline
		$h\left(  \pi\right)  =p\times p\left(  \operatorname{dit}\left(  \pi\right)
		\right)  $ & $h\left(  F:\psi\right)  =\operatorname{tr}\left[  P_{\left[
			\operatorname*{qudit}\left(  F\right)  \right]  }\rho\left(  \psi\right)
		\otimes\rho\left(  \psi\right)  \right]  $\\\hline
		$h\left(  \pi,\sigma\right)  =p\times p\left(  \operatorname{dit}\left(
		\pi\right)  \cup\operatorname{dit}\left(  \sigma\right)  \right)  $ &
		$h\left(  F,G:\psi\right)  =\operatorname{tr}\left[  P_{\left[
			\operatorname*{qudit}\left(  F\right)  \cup\operatorname*{qudit}\left(
			G\right)  \right]  }\rho\left(  \psi\right)  \otimes\rho\left(  \psi\right)
		\right]  $\\\hline
		$h\left(  \pi|\sigma\right)  =p\times p\left(  \operatorname{dit}\left(
		\pi\right)  -\operatorname{dit}\left(  \sigma\right)  \right)  $ & $h\left(
		F|G:\psi\right)  =\operatorname{tr}\left[  P_{\left[  \operatorname*{qudit}%
			\left(  F\right)  -\operatorname*{qudit}\left(  G\right)  \right]  }%
		\rho\left(  \psi\right)  \otimes\rho\left(  \psi\right)  \right]  $\\\hline
		$m\left(  \pi,\sigma\right)  =p\times p\left(  \operatorname{dit}\left(
		\pi\right)  \cap\operatorname{dit}\left(  \sigma\right)  \right)  $ &
		$m\left(  F,G:\psi\right)  =\operatorname{tr}\left[  P_{\left[
			\operatorname*{qudit}\left(  F\right)  \cap\operatorname*{qudit}\left(
			G\right)  \right]  }\rho\left(  \psi\right)  \otimes\rho\left(  \psi\right)
		\right]  $\\\hline
		$h\left(  \pi\right)  =h\left(  \pi|\sigma\right)  +m\left(  \pi
		,\sigma\right)  $ & $h\left(  F:\psi\right)  =h\left(  F|G:\psi\right)
		+m\left(  F,G:\psi\right)  $\\\hline
		$\Pr\left(  \phi_{i}\right)  =\operatorname*{tr}\left[  P_{f^{-1}\left(
			\phi_{i}\right)  }\rho\left(  f^{-1}\right)  \right]  $ & $\Pr\left(  \phi
		_{i}\right)  =\operatorname*{tr}\left[  P_{\phi_{i}}\rho\left(  \psi\right)
		\right]  $\\\hline
		$\Pr\left(  \phi_{i}\right)  =$ one-draw prob. of $\phi_{i}$ & $\Pr\left(
		\phi_{i}\right)  =$ one-meas. prob. of $\phi_{i}$\\\hline
		$h\left(  f^{-1}\right)  =$ two-draw prob. diff. $f$-values & $h\left(
		F:\psi\right)  =$ two-meas. prob. diff. $F$-eigenvalues\\\hline
		$\rho\left(  \pi\right)  =\sum_{i}P_{B_{i}}\rho\left(  \mathbf{0}_{U}\right)
		P_{B_{i}}$ & $\hat{\rho}\left(  \psi\right)  =\sum_{i}P_{V_{i}}\rho\left(
		\psi\right)  P_{V_{i}}$ (L\"{u}ders)\\\hline
		$h\left(  \pi\right)  =1-\operatorname{tr}\left[  \rho\left(  \pi\right)
		^{2}\right]  $ & $h\left(  F:\psi\right)  =1-\operatorname{tr}\left[
		\hat{\rho}\left(  \psi\right)  ^{2}\right]  $\\\hline
		$h\left(  \pi\right)  =$ sum sq. zeroed $\rho\left(  \mathbf{0}_{U}\right)
		\rightsquigarrow\rho\left(  \pi\right)  $ & $h\left(  F:\psi\right)  =$ sum
		ab. sq. zeroed $\rho\left(  \psi\right)  \rightsquigarrow\hat{\rho}\left(
		\psi\right)  $\\\hline
	\end{tabular}

	Table 10: Logical entropies with probabilities applied to ditsets and qudit spaces.
\end{center}

The last three lines of Table 10 anticipate the quantum version of the previous
results about real-valued density matrices. One of the main results about
density matrices (over the complex numbers where $\left\Vert \rho
_{ij}\right\Vert ^{2}$ is the absolute square of $\rho_{ij}$) is:

\begin{proposition}
	$\operatorname{tr}\left[  \rho^{2}\right]  =\sum_{i,j}\left\Vert \rho
	_{ij}\right\Vert ^{2}.$\cite[p. 77]{fano:density}
\end{proposition}

\textbf{Proof}: A diagonal entry in $\rho^{2}$ is $\left(  \rho^{2}\right)
_{ii}=\sum_{j=1}^{n}\rho_{ij}\rho_{ji}^{\ast}=\sum_{j=1}^{n}\left\Vert
\rho_{ij}\right\Vert ^{2}$ so $\operatorname{tr}\left[  \rho^{2}\right]
=\sum_{i=1}^{n}\left(  \rho^{2}\right)  _{ii}=\sum_{i,j}\left\Vert \rho
_{ij}\right\Vert ^{2}$. $\square$

\noindent In general the quantum logical entropy of a density matrix $\rho$
is: $h\left(  \rho\right)  =1-\operatorname{tr}\left[  \rho^{2}\right]
=1-\sum_{ij}\left\Vert \rho_{ij}\right\Vert ^{2}$. The terms $\left\Vert
\rho_{ij}\right\Vert ^{2}$ are the `indistinction' probabilities so $h\left(
\rho\right)  =1-\sum_{ij}\left\Vert \rho_{ij}\right\Vert ^{2}$ is, as in the
classical case, the sum of the probabilities of distinctions.

The change in density matrices due to a projective measurement is given by the
L\"{u}ders mixture operation. If $V_{i}$ is the eigenspace for the eigenvalue
$\phi_{i}$ of $F$ and $P_{V_{i}}$ is the projection matrix $P_{V_{i}%
}:V\rightarrow V$ to that subspace, then the post-measurement density matrix
by the L\"{u}ders mixture operation is:

\begin{center}
	$\hat{\rho}\left(  \psi\right)  =\sum_{i}P_{V_{i}}\rho\left(  \psi\right)
	P_{V_{i}}$.
\end{center}

The previous result about the sum of the squares of the non-zero off-diagonal
elements of $\rho$ that are zeroed in the transition $\rho\rightsquigarrow
\hat{\rho}$ carries over to the quantum case of density matrices over the
complex numbers.

\begin{theorem}
	[Measuring Measurement] The increase in quantum logical entropy, $h\left(  \hat{\rho}\left(
	\psi\right)  \right)  $ due to the $F$-measurement of the pure state $\psi$ is
	the sum of the absolute squares of the non-zero off-diagonal terms
	(coherences) in $\rho\left(  \psi\right)  $ (represented in a basis of
	$F$-eigenvectors) that are zeroed (decohered) in the post-measurement
	L\"{u}ders mixture density matrix $\hat{\rho}\left(  \psi\right)  =\sum
	_{i}P_{V_{i}}\rho\left(  \psi\right)  P_{V_{i}}$.
\end{theorem}

\textbf{Proof:} $h\left(  \hat{\rho}\left(  \psi\right)  \right)  -h\left(
\rho\left(  \psi\right)  \right)  =\left(  1-\operatorname*{tr}\left[
\hat{\rho}\left(  \psi\right)  ^{2}\right]  \right)  -\left(
1-\operatorname*{tr}\left[  \rho\left(  \psi\right)  ^{2}\right]  \right)
=\sum_{j,k}\left(  \left\Vert \rho_{jk}\left(  \psi\right)  \right\Vert
^{2}-\left\Vert \hat{\rho}_{jk}\left(  \psi\right)  \right\Vert ^{2}\right)  $
since $\operatorname{tr}\left[  \rho^{2}\right]  =\sum_{i,j}\left\Vert
\rho_{ij}\right\Vert ^{2}$ is the sum of the absolute squares of all the
elements of $\rho$. If $u_{j}$ and $u_{k}$ are a qudit of $F$, then and only
then are the corresponding off-diagonal terms in $\rho\left(  \psi\right)  $
zeroed by the L\"{u}ders mixture operation $\sum_{i=1}^{I}P_{V_{i}}\rho\left(
\psi\right)  P_{V_{i}}$ to obtain $\hat{\rho}\left(  \psi\right)  $ from
$\rho\left(  \psi\right)  $. $\square$

The notion of quantnum logical entropy is an example of some QM mathematics that needed to be developed to substantiate our thesis about the linearization of partition mathematics. The mere linearized definitions of quantum logical entropy prove nothing by themselves. But the definitions of density matrices, projective measurement, and the L\"{u}ders mixture operation are all standard in texts that do not consider classical or quantum logical entropy. Hence the precise connection between quantum logical entropy and projective measurement in the Measuring Measurement Theorem is an example of some QM mathematics that needed to be developed as part of our thesis about the linearized mathematics of partitions.

A careful calculation shows that $h\left(  F:\psi\right)  =h\left(  \hat{\rho
}\left(  \psi\right)  \right)  $ which also equals the sum of the absolute
squares of zeroed terms in the transition $\rho\left(  \psi\right)
\rightsquigarrow\hat{\rho}\left(  \psi\right)  $ (since pure states
$\rho\left(  \psi\right)  $ have zero quantum logical entropy). These
equalities can be illustrated by working through a simple example of measuring
$z$-axis spin.

\textbf{Example}: Let $\left\vert \psi\right\rangle =\alpha_{\uparrow
}\left\vert \uparrow\right\rangle +\alpha_{\downarrow}\left\vert
\downarrow\right\rangle =%
\begin{bmatrix}
	\alpha_{\uparrow}\\
	\alpha_{\downarrow}%
\end{bmatrix}
$ be a pure normalized superposition state of $z$-spin up and $z$-spin down so
the density matrix is $\rho\left(  \psi\right)  =%
\begin{bmatrix}
	p_{\uparrow} & \alpha_{\uparrow}\alpha_{\downarrow}^{\ast}\\
	\alpha_{\downarrow}\alpha_{\uparrow}^{\ast} & p_{\downarrow}%
\end{bmatrix}
$ (where $\alpha^{\ast}$ is the complex conjugate of $\alpha$). For the
observable $F$, let the eigenvalue function be $f:\left\{  \left\vert
\uparrow\right\rangle ,\left\vert \downarrow\right\rangle \right\}
\rightarrow\left\{  +1,-1\right\}  $ where $f(\left\vert \uparrow\right\rangle
)=1$ and $f\left(  \left\vert \downarrow\right\rangle \right)  =-1$. Then $F:%
\mathbb{C}
^{2}\rightarrow%
\mathbb{C}
^{2}$ is represented by the matrix $F=%
\begin{bmatrix}
	1 & 0\\
	0 & -1
\end{bmatrix}
$. The tensor product $\rho\left(  \psi\right)  \otimes\rho\left(
\psi\right)  $ is the $2^{2}\times2^{2}$ matrix:

\begin{center}
	$%
	\begin{bmatrix}
		p_{\uparrow}\rho\left(  \psi\right)  & \alpha_{\uparrow}\alpha_{\downarrow
		}^{\ast}\rho\left(  \psi\right) \\
		\alpha_{\downarrow}\alpha_{\uparrow}^{\ast}\rho\left(  \psi\right)  &
		p_{\downarrow}\rho\left(  \psi\right)
	\end{bmatrix}
	=%
	\begin{bmatrix}
		p_{\uparrow}^{2} & p_{\uparrow}\alpha_{\uparrow}\alpha_{\downarrow}^{\ast} &
		\alpha_{\uparrow}\alpha_{\downarrow}^{\ast}p_{\uparrow} & \alpha_{\uparrow
		}\alpha_{\downarrow}^{\ast}\alpha_{\uparrow}\alpha_{\downarrow}^{\ast}\\
		p_{\uparrow}\alpha_{\downarrow}\alpha_{\uparrow}^{\ast} & p_{\uparrow
		}p_{\downarrow} & \alpha_{\uparrow}\alpha_{\downarrow}^{\ast}\alpha
		_{\downarrow}\alpha_{\uparrow}^{\ast} & \alpha_{\uparrow}\alpha_{\downarrow
		}^{\ast}p_{\downarrow}\\
		\alpha_{\downarrow}\alpha_{\uparrow}^{\ast}p_{\uparrow} & \alpha_{\downarrow
		}\alpha_{\uparrow}^{\ast}\alpha_{\uparrow}\alpha_{\downarrow}^{\ast} &
		p_{\downarrow}p_{\uparrow} & p_{\downarrow}\alpha_{\uparrow}\alpha
		_{\downarrow}^{\ast}\\
		\alpha_{\downarrow}\alpha_{\uparrow}^{\ast}\alpha_{\downarrow}\alpha
		_{\uparrow}^{\ast} & \alpha_{\downarrow}\alpha_{\uparrow}^{\ast}p_{\downarrow}
		& p_{\downarrow}\alpha_{\downarrow}\alpha_{\uparrow}^{\ast} & p_{\downarrow
		}^{2}%
	\end{bmatrix}
	$.
\end{center}

\noindent The qudits of $F$ are $\left\vert \uparrow\right\rangle
\otimes\left\vert \downarrow\right\rangle $ and $\left\vert \downarrow
\right\rangle \otimes\left\vert \uparrow\right\rangle $ so the projection
matrix to the subspace $\left[  \operatorname*{qudit}\left(  F\right)
\right]  $ of $%
\mathbb{C}
^{2}\otimes%
\mathbb{C}
^{2}$ generated by those qudits is:

\begin{center}
	$%
	\begin{bmatrix}
		0 & 0 & 0 & 0\\
		0 & 1 & 0 & 0\\
		0 & 0 & 1 & 0\\
		0 & 0 & 0 & 0
	\end{bmatrix}
	$.
\end{center}

\noindent Hence the quantum logical entropy $h\left(  F:\psi\right)
=\operatorname{tr}\left[  P_{\left[  qudit\left(  F\right)  \right]  }%
\rho\left(  \psi\right)  \otimes\rho\left(  \psi\right)  \right]  $ is:

\begin{center}
	$h\left(  F:\psi\right)  =\operatorname*{tr}\left(
	\begin{bmatrix}
		0 & 0 & 0 & 0\\
		0 & 1 & 0 & 0\\
		0 & 0 & 1 & 0\\
		0 & 0 & 0 & 0
	\end{bmatrix}%
	\begin{bmatrix}
		p_{\uparrow}^{2} & p_{\uparrow}\alpha_{\uparrow}\alpha_{\downarrow}^{\ast} &
		\alpha_{\uparrow}\alpha_{\downarrow}^{\ast}p_{\uparrow} & \alpha_{\uparrow
		}\alpha_{\downarrow}^{\ast}\alpha_{\uparrow}\alpha_{\downarrow}^{\ast}\\
		p_{\uparrow}\alpha_{\downarrow}\alpha_{\uparrow}^{\ast} & p_{\uparrow
		}p_{\downarrow} & \alpha_{\uparrow}\alpha_{\downarrow}^{\ast}\alpha
		_{\downarrow}\alpha_{\uparrow}^{\ast} & \alpha_{\uparrow}\alpha_{\downarrow
		}^{\ast}p_{\downarrow}\\
		\alpha_{\downarrow}\alpha_{\uparrow}^{\ast}p_{\uparrow} & \alpha_{\downarrow
		}\alpha_{\uparrow}^{\ast}\alpha_{\uparrow}\alpha_{\downarrow}^{\ast} &
		p_{\downarrow}p_{\uparrow} & p_{\downarrow}\alpha_{\uparrow}\alpha
		_{\downarrow}^{\ast}\\
		\alpha_{\downarrow}\alpha_{\uparrow}^{\ast}\alpha_{\downarrow}\alpha
		_{\uparrow}^{\ast} & \alpha_{\downarrow}\alpha_{\uparrow}^{\ast}p_{\downarrow}
		& p_{\downarrow}\alpha_{\downarrow}\alpha_{\uparrow}^{\ast} & p_{\downarrow
		}^{2}%
	\end{bmatrix}
	\right)  $
	
	$=\operatorname*{tr}\left(  \allowbreak%
	\begin{bmatrix}
		0 & 0 & 0 & 0\\
		\alpha_{\downarrow}p_{\uparrow}\alpha_{\uparrow}^{\ast} & p_{\uparrow
		}p_{\downarrow} & \alpha_{\uparrow}\alpha_{\downarrow}\alpha_{\uparrow}^{\ast
		}\alpha_{\downarrow}^{\ast} & \alpha_{\uparrow}p_{\downarrow}\alpha
		_{\downarrow}^{\ast}\\
		\alpha_{\downarrow}p_{\uparrow}\alpha_{\uparrow}^{\ast} & \alpha_{\uparrow
		}\alpha_{\downarrow}\alpha_{\uparrow}^{\ast}\alpha_{\downarrow}^{\ast} &
		p_{\uparrow}p_{\downarrow} & \alpha_{\uparrow}p_{\downarrow}\alpha
		_{\downarrow}^{\ast}\\
		0 & 0 & 0 & 0
	\end{bmatrix}
	\right)  =2p_{\uparrow}p_{\downarrow}$.
\end{center}

The second way to calculate the quantum logical entropy of the
post-measurement state is using the L\"{u}ders mixture operation. The
measurement of that spin-observable $F$ goes from the pure state $\rho\left(
\psi\right)  $ to%
\begin{align}
	&  P_{\uparrow}\rho\left(  \psi\right)  P_{\uparrow}+P_{\downarrow}\rho\left(
	\psi\right)  P_{\downarrow}\nonumber\\
	&  =%
	\begin{bmatrix}
		1 & 0\\
		0 & 0
	\end{bmatrix}%
	\begin{bmatrix}
		p_{\uparrow} & \alpha_{\uparrow}\alpha_{\downarrow}^{\ast}\\
		\alpha_{\downarrow}\alpha_{\uparrow}^{\ast} & p_{\downarrow}%
	\end{bmatrix}%
	\begin{bmatrix}
		1 & 0\\
		0 & 0
	\end{bmatrix}
	+%
	\begin{bmatrix}
		0 & 0\\
		0 & 1
	\end{bmatrix}%
	\begin{bmatrix}
		p_{\uparrow} & \alpha_{\uparrow}\alpha_{\downarrow}^{\ast}\\
		\alpha_{\downarrow}\alpha_{\uparrow}^{\ast} & p_{\downarrow}%
	\end{bmatrix}%
	\begin{bmatrix}
		0 & 0\\
		0 & 1
	\end{bmatrix}
	\nonumber\\
	&  =%
	\begin{bmatrix}
		p_{\uparrow} & 0\\
		0 & p_{\downarrow}%
	\end{bmatrix}
	=\hat{\rho}\left(  \psi\right)  .\nonumber
\end{align}

\noindent The logical entropy of $\hat{\rho}\left(  \psi\right)  $ is:

\begin{center}
	$h\left(  \hat{\rho}\left(  \psi\right)  \right)  =1-\operatorname{tr}\left[
	\hat{\rho}\left(  \psi\right)  ^{2}\right]  =1-p_{\uparrow}^{2}-p_{\downarrow
	}^{2}=2p_{\uparrow}p_{\downarrow}=h\left(  F:\psi\right)  $.
\end{center}

\noindent The third way to calculate the quantum logical entropy of $\hat
{\rho}\left(  \psi\right)  $ is to sum the absolute squares of the non-zero
off-diagonal terms in the pure state density matrix $\rho\left(  \psi\right)
$ that are zeroed in the transition to the post-measurement density matrix
$\hat{\rho}\left(  \psi\right)  $, i.e.,

\begin{center}
	$\rho\left(  \psi\right)  =%
	\begin{bmatrix}
		p_{\uparrow} & \alpha_{\uparrow}\alpha_{\downarrow}^{\ast}\\
		\alpha_{\downarrow}\alpha_{\uparrow}^{\ast} & p_{\downarrow}%
	\end{bmatrix}
	\rightsquigarrow%
	\begin{bmatrix}
		p_{\uparrow} & 0\\
		0 & p_{\downarrow}%
	\end{bmatrix}
	=\hat{\rho}\left(  \psi\right)  $,
\end{center}

\noindent and that sum is:

\begin{center}
	$2\alpha_{\uparrow}\alpha_{\downarrow}^{\ast}\alpha_{\downarrow}%
	\alpha_{\uparrow}^{\ast}=2p_{\uparrow}p_{\downarrow}=h\left(  \hat{\rho
	}\left(  \psi\right)  \right)  =$ $h\left(  F:\psi\right)  $.
\end{center}

\noindent It might be noted that the Born rule is built into the density
matrix formulation, e.g.,

\begin{center}
	$\Pr\left(  \left\vert \uparrow\right\rangle \right)  =\operatorname*{tr}%
	\left[  P_{\uparrow}\rho\left(  \psi\right)  \right]  =\operatorname*{tr}%
	\left(
	\begin{bmatrix}
		1 & 0\\
		0 & 0
	\end{bmatrix}%
	\begin{bmatrix}
		p_{\uparrow} & \alpha_{\uparrow}\alpha_{\downarrow}^{\ast}\\
		\alpha_{\downarrow}\alpha_{\uparrow}^{\ast} & p_{\downarrow}%
	\end{bmatrix}
	\right)  =\alpha_{\uparrow}\alpha_{\uparrow}^{\ast}=p_{\uparrow}$.
\end{center}

Quantum measurement creates distinctions, e.g., the distinction between
spin-up and spin-down in the spin example, and quantum logical entropy
precisely measures those distinctions. We started `classically' with a
universe set of distinct elements

\begin{quotation}
	\noindent The set can be thought of as being originally fully distinct, while
	each partition collects together blocks whose distinctions are factored out.
	Each block represents elements that are associated with an equivalence
	relation on the set. Then, the elements of a block are indistinct among
	themselves while different blocks are distinct from each other, given an
	equivalence relation. \cite[p. 1]{tamir:4open}
\end{quotation}

Then the Yoga of Linearization translated the concepts of
information-as-distinctions to the corresponding vector space concept which
would be the concepts of quantum information-as-qudits. One of the founders of
quantum information theory, Charles Bennett, captured the target concept of
information as distinctions, differences, and distinguishability: ``So
information really is a very useful abstraction. It is the notion of
distinguishability abstracted away from what we are distinguishing, or from
the carrier of information...".\cite[p. 155]{bennett:qinfo}

\begin{quotation}
	With these concepts in mind, it seems that the extension of this framework of
	partitions and distinctions to the study of quantum systems may bring new
	insights into problems of quantum state discrimination, quantum cryptography,
	and quantum channel capacity. In fact, in these problems, we are, in one way
	or another, interested in a distance measure between distinguishable states,
	which is exactly the kind of knowledge the logical entropy is associated with.
	\cite[p. 1]{tamir:4open}
\end{quotation}

\subsection{Group representation theory}

Group representation theory is a key part of the mathematics of QM. This
immediately supports our thesis since a group representation is essentially a
`dynamic' or `active' way to define an equivalence relation; each group
operation transforms an element into an equivalent or symmetric element.

Given a \textit{set} $G$ indexing (associative) mappings $\left\{
R_{g}:U\rightarrow U\right\}  _{g\in G}$ on a set $U$, what are the conditions
on the set of  mappings so that it is a set representation of a group? Define
the binary relation $R$ on $U\times U$ by:

\begin{center}
$\left(  u,u^{\prime}\right)  \in R$ if $\exists g\in G$ such that
$R_{g}\left(  u\right)  =u^{\prime}$.
\end{center}

\noindent Then the conditions that make $R_{g}$ into a group representation
are the conditions that imply $R$ is an equivalence relation:

\begin{enumerate}
\item existence of the identity $1_{U}\in G$ implies reflexivity of $R$;

\item existence of inverses implies symmetry of $R$; and

\item closure under products, i.e., for $g,g^{\prime}\in G$, $\exists
g^{\prime\prime}\in G$ such that $R_{g^{\prime\prime}}=R_{g^{\prime}}R_{g}$,
implies transitivity of $R$.
\end{enumerate}

Thus a set representation $\left\{  R_{g}\right\}  _{g\in G}$ of a group $G$
(or group action on a set) is essentially a `dynamic' way to define an
equivalence relation $R$ on the set \cite{castellani:symeq}. The minimal
invariant (or irreducible) subsets of the set representation are the
\textit{orbits}, and they are the equivalence classes of the equivalence
relation $R$ or blocks of the orbit partition. The restriction of a set
representation of a group to an orbit is an \textit{irreducible
representation} or \textit{irrep}.

Let $f:U\rightarrow%
\mathbb{R}
$ be a numerical attribute on $U$ and consider the $n\times n$ diagonal matrix
$\tilde{f}:%
\mathbb{R}
^{n}\rightarrow%
\mathbb{R}
^{n}$ with the diagonal entries $\left(  \tilde{f}\right)  _{ii}=f\left(
u_{i}\right)  $. Let $M:%
\mathbb{R}
^{n}\rightarrow%
\mathbb{R}
^{n}$ be any matrix that commutes with $\tilde{f}$, $\tilde{f}M=M\tilde{f}$,
i.e., the following diagram commutes:

\begin{center}
$%
\begin{array}
[c]{ccc}%
\mathbb{R}
^{n} & \overset{M}{\longrightarrow} &
\mathbb{R}
^{n}\\
\tilde{f}\downarrow^{{}} &  & ^{{}}\downarrow\tilde{f}\\%
\mathbb{R}
^{n} & \overset{M}{\longrightarrow} &
\mathbb{R}
^{n}%
\end{array}
$.
\end{center}

\noindent Then computing the $ik$ entries: $\left(  \tilde{f}M\right)
_{ik}=f\left(  u_{i}\right)  M_{ik}=M_{ik}f\left(  u_{k}\right)  =\left(
M\tilde{f}\right)  _{ik}$. Thus $\left[  f\left(  u_{i}\right)  -f\left(
u_{k}\right)  \right]  M_{ik}=0$ so if $\left(  u_{i},u_{k}\right)  $ is a dit
of the partition $f^{-1}$, i.e., $f\left(  u_{i}\right)  \neq f\left(
u_{k}\right)  $, then $M_{ik}=0$.\footnote{In the general vector space case,
for two commuting (diagonalizable) operators, $FG=GF$, if both are represented
in a basis of $F$-eigenvectors and $\left(  u_{i},u_{k}\right)  $ is a qudit
of $F$, then $G_{ik}=0$. \cite[p. 4]{tinkham:gt-qm}}

A numerical attribute $f:U\rightarrow%
\mathbb{R}
$ is said to \textit{commute} with the set representation $R=\left\{
R_{g}\right\}  _{g\in G}$ if for any $R_{g}$, the following diagram commutes:

\begin{center}
$%
\begin{array}
[c]{ccc}%
U & \overset{R_{g}}{\longrightarrow} & U\\
& \searrow^{f} & ^{{}}\downarrow^{f}\\
&  &
\mathbb{R}%
\end{array}
$.
\end{center}

\noindent Taking $R_{g}:%
\mathbb{R}
^{n}\rightarrow%
\mathbb{R}
^{n}$ as a permutation matrix, then $f$ commuting with $R=\left\{
R_{g}\right\}  _{g\in G}$ means $\tilde{f}R_{g}=R_{g}\tilde{f}$ so if $\left(
u_{i},u_{k}\right)  \in\operatorname*{dit}\left(  f^{-1}\right)  $, then
$\left(  R_{g}\right)  _{ik}=0$ for all $g\in G$. A subset $S\subseteq U$ is
said to be \textit{invariant} under $R$ if $R_{g}\left(  S\right)  \subseteq
S$ for all $g\in G$. The blocks $f^{-1}\left(  r\right)  $ for $r\in f\left(
U\right)  $ for a commuting $f$ are invariant subsets of $U$ under $R$. Thus
the partition $f^{-1}$ for a commuting $f$ is refined by the orbit partition
since orbits are the \textit{minimal} invariant subsets. A set of commuting
attributes $f$, $g$,..., $h$ is said to be \textit{complete} if the join
$f^{-1}\vee g^{-1}\vee...\vee h^{-1}$ is the orbit partition.

Given a finite group $G$ and a finite-dimensional vector space $V$ over $%
\mathbb{C}
$, a \textit{vector space representation} of the group $G$ is a map
$R:G\rightarrow GL\left(  V\right)  $ (the general linear group of invertible
linear maps on $V$) where $g\longmapsto R_{g}:V\rightarrow V$ from $G$ to
invertible linear maps on $V$ such that $R_{1}=I$ and $R_{g^{\prime}}%
R_{g}=R_{g^{\prime}g}$. Using the Yoga to lift set concepts to vector space
concepts, the notion of a minimal invariant subset or orbit yields the notion
of a minimal invariant subspace which is called an \textit{irreducible
subspace} of $V$. Just as the orbits of a set representation form a partition
of $U$, so the irreducible subspaces of a vector space representation form a
direct-sum decomposition of $V$. The restriction of a vector space
representation of a group to an irreducible subspace is an \textit{irreducible
representation} or \textit{irrep}.

Let $\left\{  R_{g}\right\}  _{g\in G}$ be a set representation of $G$ on a
set $U$ that is an ON basis set for $V$ and let $f:U\rightarrow%
\mathbb{R}
$ be a commuting attribute. Then for any $u_{i}\in U$ and any $g\in G$, if
$R_{g}\left(  u_{i}\right)  =u_{k}\in U$, then by commutativity, $f\left(
u_{i}\right)  =$ $f\left(  R_{g}\left(  u_{i}\right)  \right)  =f\left(
u_{k}\right)  $. Then a unitary operator $F:V\rightarrow V$ is defined by
$Fu=f\left(  u\right)  u$ and the group representation $R_{g}$ extends to $V$
from its definition on the basis set $U$. Does the operator $F$ commute with
$R_{g}$ in the sense that for each $R_{g}$, the following diagram commutes?

\begin{center}
$%
\begin{array}
[c]{ccc}%
V & \overset{R_{g}}{\longrightarrow} & V\\
^{F}\downarrow~ &  & ^{{}}\downarrow^{F}\\
V & \overset{R_{g}}{\longrightarrow} & V
\end{array}
$.
\end{center}

\noindent Starting with a basis element $u_{i}\in V$ and going around the
square clockwise, $R_{g}\left(  u_{i}\right)  =u_{k}\in V$ is taken to
$Fu_{k}=f\left(  u_{k}\right)  u_{k}\in V$. Going around the square
counterclockwise, $Fu_{i}=f\left(  u_{i}\right)  u_{i}$ and $R_{g}\left(
f\left(  u_{i}\right)  u_{i}\right)  =f\left(  u_{i}\right)  u_{k}=f\left(
u_{k}\right)  u_{k}$ so the square commutes. Hence the set-concept of a
commuting $f$ extends by the Yoga to the usual concept of an observable $F$
commuting with a vector space representation.\footnote{The Yoga is used to
generate vector space concepts corresponding to set concepts. There is no
implication that every instance of a vector space concept, e.g., a commuting
$F$, must come from an instance of the set concept, e.g., a commuting $f$..}

The eigenspaces of a commuting $F$ are invariant under $R$. Then the DSD of
eigenspaces for a commuting $F$ is refined (defined in the obvious way
\cite{ell:dsds}) by the DSD of irreducible subspaces. A set of operators
commuting with $R$ such as $F$, $G$,..., $H$ is said to be \textit{complete}
if the join-like operation on their DSDs has all its subspaces as irreducible.

\textbf{Schur's Lemma} (set case): A commuting $f$ restricted to (i.e.,
$f\upharpoonright$) an irreducible subset (i.e., an orbit) is a constant function.

\textbf{Schur's Lemma} (vector space case): A commuting $F$ restricted to
(i.e., $F\upharpoonright$) an irreducible subspace is a constant operator.

The Yoga of Linearization applied to group representations is illustrated in
Table 11.

\begin{center}%
\begin{tabular}
[c]{|l|l|l|}\hline
Yoga & Set representations & Vector space reps.\\\hline\hline
Representation & $\left\{  R_{g}:U\rightarrow U\right\}  _{g\in G}$ &
$\left\{  R_{g}:V\rightarrow V\right\}  _{g\in G}$\\\hline
Min. Invariants & Orbits & Irreducible subspaces\\\hline
Partition & Orbit partition & DSD of irred. subspaces\\\hline
Irreps & Rep. on orbits & Rep. on irred. subspaces\\\hline
Commuting & $f:U\rightarrow%
\mathbb{R}
$, $\forall g$, $fR_{g}=f$ & $F:V\rightarrow V$, $\forall g$, $FR_{g}=R_{g}%
F$\\\hline
Invariants & $f^{-1}\left(  r\right)  $ commuting $f$ & Eigenspaces commuting
$F$\\\hline
Schur's Lemma & Comm. $f\upharpoonright$orbit const. & Comm. $F\upharpoonright
$irred. sp. constant\\\hline
\end{tabular}

Table 11: Summary of Yoga of Linearization for group representations.
\end{center}

\textbf{Example}: Consider the Klein four-group written additively: $G=%
\mathbb{Z}
_{2}\times%
\mathbb{Z}
_{2}=\left\{  \left(  0,0\right)  ,\left(  1,0\right)  ,\left(  0,1\right)
,\left(  1,1\right)  \right\}  $. The \textit{Cayley group space} of that
group is the complex vector space $\left\{
\mathbb{Z}
_{2}\times%
\mathbb{Z}
_{2}\rightarrow%
\mathbb{C}
\right\}  $ of all complex-valued maps on the four-element set $%
\mathbb{Z}
_{2}\times%
\mathbb{Z}
_{2}$. A basis for the four-dimensional space $%
\mathbb{C}
^{4}$ is the set of maps $\left\vert g^{\prime}\right\rangle $ which take
value $1$ on $g^{\prime}$ and $0$ on the other $g\in G$. The action of the
group on this space is defined by $R_{g}\left(  \left\vert g^{\prime
}\right\rangle \right)  =\left\vert g+g^{\prime}\right\rangle $. The group
action just permutes the basis vectors in the Cayley group space and would be
represented by permutation matrices. For the ordering $\left(  0,0\right)
,\left(  1,0\right)  ,\left(  0,1\right)  ,$ and $\left(  1,1\right)  $ on $%
\mathbb{Z}
_{2}\times%
\mathbb{Z}
_{2}$, the non-identity permutation operators have the matrices;

\begin{center}
$R_{\left(  1,0\right)  }=%
\begin{bmatrix}
0 & 1 & 0 & 0\\
1 & 0 & 0 & 0\\
0 & 0 & 0 & 1\\
0 & 0 & 1 & 0
\end{bmatrix}
$; $R_{\left(  0,1\right)  }=%
\begin{bmatrix}
0 & 0 & 1 & 0\\
0 & 0 & 0 & 1\\
1 & 0 & 0 & 0\\
0 & 1 & 0 & 0
\end{bmatrix}
$; $R_{\left(  1,1\right)  }=%
\begin{bmatrix}
0 & 0 & 0 & 1\\
0 & 0 & 1 & 0\\
0 & 1 & 0 & 0\\
1 & 0 & 0 & 0
\end{bmatrix}
$.
\end{center}

\noindent The group is Abelian, so each of these operators can be viewed as an
observable that commutes with the $R_{g}$ for $g\in G$, and then its
eigenspaces will be invariant under the group operations.

For $R_{\left(  1,0\right)  }$, the invariant eigenspaces with their
eigenvalues and generating eigenvectors are:

\begin{center}
$\left\{
\begin{bmatrix}
1\\
-1\\
1\\
-1
\end{bmatrix}
,%
\begin{bmatrix}
1\\
-1\\
-1\\
1
\end{bmatrix}
\right\}  \leftrightarrow\lambda=-1,\left\{
\begin{bmatrix}
1\\
1\\
1\\
1
\end{bmatrix}
,%
\begin{bmatrix}
1\\
1\\
-1\\
-1
\end{bmatrix}
\right\}  \leftrightarrow\lambda=1\allowbreak$.
\end{center}

For $R_{\left(  0,1\right)  }$, they are:

\begin{center}
$\left\{
\begin{bmatrix}
1\\
-1\\
-1\\
1
\end{bmatrix}
,%
\begin{bmatrix}
1\\
1\\
-1\\
-1
\end{bmatrix}
\right\}  \leftrightarrow\lambda=-1,\left\{
\begin{bmatrix}
1\\
1\\
1\\
1
\end{bmatrix}
,%
\begin{bmatrix}
1\\
-1\\
1\\
-1
\end{bmatrix}
\right\}  \leftrightarrow\lambda=1\allowbreak$.
\end{center}

Since the two operators commute, their eigenspace DSDs commute so we can take
their join. The blocks of the join are a DSD and are automatically invariant.
Since the blocks of the join are one-dimensional, those four subspaces are
also irreducible and thus the two operators form a complete set of commuting
operators (CSCO). The commuting operators always have a set of simultaneous
eigenvectors, and we have arranged the generating eigenvectors of the
eigenspaces so that they are all simultaneous eigenvectors which can, as
usual, be characterized by kets using the respective eigenvalues;

\begin{center}
$%
\begin{bmatrix}
1\\
1\\
1\\
1
\end{bmatrix}
=\left\vert 1,1\right\rangle $; $%
\begin{bmatrix}
1\\
-1\\
1\\
-1
\end{bmatrix}
=\left\vert -1,1\right\rangle $; $%
\begin{bmatrix}
1\\
1\\
-1\\
-1
\end{bmatrix}
=\left\vert 1,-1\right\rangle $; $%
\begin{bmatrix}
1\\
-1\\
-1\\
1
\end{bmatrix}
=\left\vert -1,-1\right\rangle $.
\end{center}

The restrictions of the group representation to these four irreducible
subspaces gives the four irreducible representations or irreps of the group.
Since any vector can be uniquely decomposed into the sum of vectors in the
irreducible subspaces, the representation on the whole space can be expressed,
in the obvious sense, as the direct sum of the irreps.

Moving to non-Abelian groups, not to mention Lie groups, greatly increases the
mathematical complexity. But for our purposes, the point is that the key
concepts of group representations for QM come out of the partitional
mathematics of definiteness and indefiniteness. The irreps give all the
different ways that minimal definite alternatives are defined consistent with
the indistinction-creating symmetries of the group. The properties of the
eigen-alternatives are determined by the irreps of the symmetry group of the
Hamiltonian or as the elementary particles themselves are determined by the
irreps of groups in particle physics. The irreps fill out the symmetry-adapted possibilities.

Since the group action creates indistinctions as symmetries, moving to the
action of a subgroup means less indistinctions and more distinctions, i.e.,
symmetry-breaking, which is the second method, in addition to the join-like
operations, to move from an indefinite state to a more definite state, e.g.,
in the description of the Big Bang \cite{pagels:persymm}.

Classical physics also has symmetries so groups will play an important role,
e.g., the Noether Theorems. However group representations play a more
fundamental role in quantum mechanics as might be expected from their role in
the mathematics of partitions. Partly, this is due to the objectively
indefinite states given by the superposition principle. At a more fundamental
level, it is due to the irreps (the restriction of representations to the
irreducible subspaces) of certain group representations defining the
elementary particles at the quantum level.

\begin{quotation}
The reason is fundamentally, that the variety of states is much greater in
quantum theory than in classical physics and that there is, on the other hand,
the principle of superposition to provide a structure for the greatly
increased manifold of quantum mechanical states. The principle of
superposition renders possible the definition of the states the transformation
properties of which are particularly simple. It can in fact be shown that
every state of any quantum mechanical system, no matter what type of
interactions are present, can be considered as a superposition of states of
elementary systems. The elementary systems correspond mathematically to
irreducible representations of the Lorentz group and as such can be
enumerated. \cite[p. 8]{wigner:invar}
\end{quotation}

Prior to the discussion of group representation theory, we started with a
\textit{given} set $U$ of distinct elements which were then taken as a given
basis set for a vector space. But those elementary given distinct elements or
basis vectors are \textit{not} given in group representation theory; they are
instead determined by the group as the irreducible subspaces and irreducible
representations (irreps) of the representation. The irreps are the elementary
symmetry-adapted eigen-alternatives determined by the group of transformations.

For a certain symmetry group of particle physics, ``an elementary particle `is'
an irreducible unitary representation of the group." \cite[p. 149]%
{stern:group} Thus our partitional approach comports with ``the soundness of
programs that ground particle properties in the irreducible representations of
symmetry transformations..." \cite[p. 171]{fine:shaky} These alternatives are
carved out by the joins of the vector space partitions of CSCOs--which
constitute a ``systematic theory ... established for the rep group based on
Dirac's CSCO (complete set of commuting operators) approach in quantum
mechanics" (\cite[p. 211]{chen:1985}, also \cite{chen:text};
\cite{wang:newapp}).

This all goes back to the transformations of group representations being
`dynamic' or `active' ways to define partitions (equivalence relations) and
their vector space versions (DSDs).

\section{Concluding remarks}

We have taken the mathematics of QM to be sufficiently represented by:

\begin{itemize}

\item commuting, non-commuting, and conjugate observables;

\item evolution by the Schr\"{o}dinger equation, the vN Type 2 process;

\item measurement and the collapse postulate;

\item quantum statistics for indistinguishable particles; 

\item projective measurement and L\"{u}ders mixture operation, the von Neumann
(vN) Type I process;and

\item group representation theory applied to quantum mechanics.
\end{itemize}

\noindent And we have argued, in each case, that the mathematics of QM is the
linearized to (Hilbert) vector space version of the mathematics of partitions.
This is not just a coincidence or an accident. The mathematics tell us
something about the unintuitive reality that QM so successfully describes.

\begin{itemize}
\item The key analytical concepts were the notions of definiteness versus
indefiniteness, distinctions versus indistinctions, and distinguishability
versus indistinguishability.

\item The key machinery for moving from indefinite to more definite states was
the partition-join-like operation of projective measurement--which is also
quantitatively measured by quantum logical entropy (the vector space version
of logical entropy at the set level).

\item To arrive at a maximally definite state determination of a CSCO, the key
operation was the partition-join operation on commuting DSDs (vector space partitions).

\end{itemize}

The `standard' reality-oriented interpretations of QM (e.g., Bohmian
mechanics, spontaneous collapse, or many-worlds) make little or no use of
those key concepts and machinery. Our approach of showing the set-level
origins of the mathematics of QM takes the formalism as being
complete--without any addition of other variables, other equations, or
other-worldly interpretations of distinction-preserving (unitary) evolution
and distinction-creating measurement. The reality-agnostic Copenhagen
interpretation also takes the formalism of QM as being complete, so the
partition mathematics approach could also be viewed as specifying the key
concepts and machinery as well as supplying a dash of realism in the form of
simplified images of properties and processes at the quantum level.

Objective indefiniteness at the quantum level violates our common-sense
classical assumption of reality as being definite all the way down. The usual
imagery of a superposition as the combination of two definite states (or
waves) to yield another definite state (or wave) needs to be replaced by an
indefiniteness imagery. The superposition of definite (or eigen) states (e.g.,
the qubit in quantum computing) should be seen as a state that is indefinite
on the differences between the superposed states--which can be better
understood if one takes the notion (and the underlying partitional math) of
indefiniteness seriously.

\begin{quotation}
These statements ... may collectively be called ``the Literal Interpretation"
of quantum mechanics. This is the interpretation resulting from taking the
formalism of quantum mechanics literally, as giving a representation of
physical properties themselves, rather than of human knowledge of them, and by
taking this representation to be complete. \cite[pp. 6-7]{shim:vienna}
\end{quotation}

\noindent The ``Follow the Math!" approach shows that the math of QM comes from
the math of partitions, and that picks out the key analytical concepts and
machinery--and also allows some imagery, albeit simplified, of the nature of
the physical properties and processes at the quantum level.

\textbf{Statements and Declarations}

There are no competing interests, no sources of funding, and no acknowledgments.

\end{document}